\documentclass[aps,prl,twocolumn,showkeys,superscriptaddress]{revtex4-1}

\usepackage[artemisia]{textgreek}
\usepackage{amssymb}
\usepackage{epsfig}
\usepackage{amsmath}
\usepackage{graphicx}
\usepackage{dcolumn}
\usepackage{latexsym}
\usepackage{color}
\usepackage{epstopdf}
\usepackage{subfigure}
\usepackage{comment}
\usepackage{nicefrac}
\usepackage{blindtext}
\usepackage[ulem=normalem]{changes}
\usepackage{float}
\usepackage{xcolor}
\usepackage{soul}
\usepackage[T1]{fontenc}
\usepackage[hidelinks]{hyperref}
\usepackage{xcolor}
\setcitestyle{super}
\hypersetup{
    colorlinks,
    linkcolor={red!50!black},
    citecolor={blue!50!black},
    urlcolor={blue!80!black}
}

\newcommand*{\citen}[1]{%
  \begingroup
    \romannumeral-`\x 
    \setcitestyle{numbers}%
    \cite{#1}%
  \endgroup   
}

\begin{document}


\title{Transparent Gatable Superconducting Shadow Junctions}

\author{Sabbir A. Khan}
\email{sabbir.khan@nbi.ku.dk}
\affiliation{Microsoft Quantum Materials Lab Copenhagen, 2800 Lyngby, Denmark}
\affiliation{Center for Quantum Devices, Niels Bohr Institute, University of Copenhagen, 2100 Copenhagen, Denmark}

\author{Charalampos Lampadaris}
\affiliation{Microsoft Quantum Materials Lab Copenhagen, 2800 Lyngby, Denmark}
\affiliation{Center for Quantum Devices, Niels Bohr Institute, University of Copenhagen, 2100 Copenhagen, Denmark}

\author{Ajuan Cui}
\affiliation{Microsoft Quantum Materials Lab Copenhagen, 2800 Lyngby, Denmark}
\affiliation{Center for Quantum Devices, Niels Bohr Institute, University of Copenhagen, 2100 Copenhagen, Denmark}

\author{Lukas Stampfer}
\affiliation{Center for Quantum Devices, Niels Bohr Institute, University of Copenhagen, 2100 Copenhagen, Denmark}

\author{Yu Liu}
\affiliation{Microsoft Quantum Materials Lab Copenhagen, 2800 Lyngby, Denmark}
\affiliation{Center for Quantum Devices, Niels Bohr Institute, University of Copenhagen, 2100 Copenhagen, Denmark}

\author{S. J. Pauka}
\affiliation{Microsoft Quantum Sydney, The University of Sydney, Sydney, NSW 2006, Australia}

\author{Martin E. Cachaza}
\affiliation{Microsoft Quantum Materials Lab Copenhagen, 2800 Lyngby, Denmark}
\affiliation{Center for Quantum Devices, Niels Bohr Institute, University of Copenhagen, 2100 Copenhagen, Denmark}

\author{Elisabetta M. Fiordaliso}
\affiliation{DTU Nanolab, Technical University of Denmark, 2800 Kgs. Lyngby, Denmark}

\author{Jung-Hyun Kang}
\affiliation{Microsoft Quantum Materials Lab Copenhagen, 2800 Lyngby, Denmark}
\affiliation{Center for Quantum Devices, Niels Bohr Institute, University of Copenhagen, 2100 Copenhagen, Denmark}

\author{Svetlana Korneychuk}
\affiliation{QuTech and Kavli Institute of Nanoscience, Delft University of Technology, 2600 GA Delft, The Netherlands}

\author{Timo Mutas}
\affiliation{Center for Quantum Devices, Niels Bohr Institute, University of Copenhagen, 2100 Copenhagen, Denmark}

\author{Joachim E. Sestoft}
\affiliation{Center for Quantum Devices, Niels Bohr Institute, University of Copenhagen, 2100 Copenhagen, Denmark}

\author{Filip Krizek}
\affiliation{Center for Quantum Devices, Niels Bohr Institute, University of Copenhagen, 2100 Copenhagen, Denmark}

\author{Rawa Tanta}
\affiliation{Microsoft Quantum Materials Lab Copenhagen, 2800 Lyngby, Denmark}
\affiliation{Center for Quantum Devices, Niels Bohr Institute, University of Copenhagen, 2100 Copenhagen, Denmark}

\author{M. C. Cassidy}
\affiliation{Microsoft Quantum Sydney, The University of Sydney, Sydney, NSW 2006, Australia}

\author{Thomas S. Jespersen}
\affiliation{Center for Quantum Devices, Niels Bohr Institute, University of Copenhagen, 2100 Copenhagen, Denmark}

\author{Peter Krogstrup}
\email{krogstrup@nbi.dk}
\affiliation{Microsoft Quantum Materials Lab Copenhagen, 2800 Lyngby, Denmark}
\affiliation{Center for Quantum Devices, Niels Bohr Institute, University of Copenhagen, 2100 Copenhagen, Denmark}

\date{\today}

\begin{abstract}
 
Gate tunable junctions are key elements in quantum devices based on hybrid semiconductor-superconductor materials. They serve multiple purposes ranging from tunnel spectroscopy probes to voltage-controlled qubit operations in gatemon and topological qubits. Common to all is that junction transparency plays a critical role. In this study, we grow single crystalline InAs, InSb and $\mathrm{InAs_{1-x}Sb_x}$ nanowires with epitaxial superconductors and \textit{in-situ} shadowed junctions in a single-step molecular beam epitaxy process. We investigate correlations between fabrication parameters, junction morphologies, and electronic transport properties of the junctions and show that the examined \textit{in-situ} shadowed junctions are of significantly higher quality than the etched junctions. By varying the edge sharpness of the shadow junctions we show that the sharpest edges yield the highest junction transparency for all three examined semiconductors. Further, critical supercurrent measurements reveal an extraordinarily high $I_\mathrm{C} R_\mathrm{N}$, close to the KO$-$2 limit. This study demonstrates a promising engineering path towards reliable gate-tunable superconducting qubits.

\end{abstract}

\maketitle


\begin{figure*}[ht!]
\vspace{0.2cm}
\includegraphics[scale=.519]{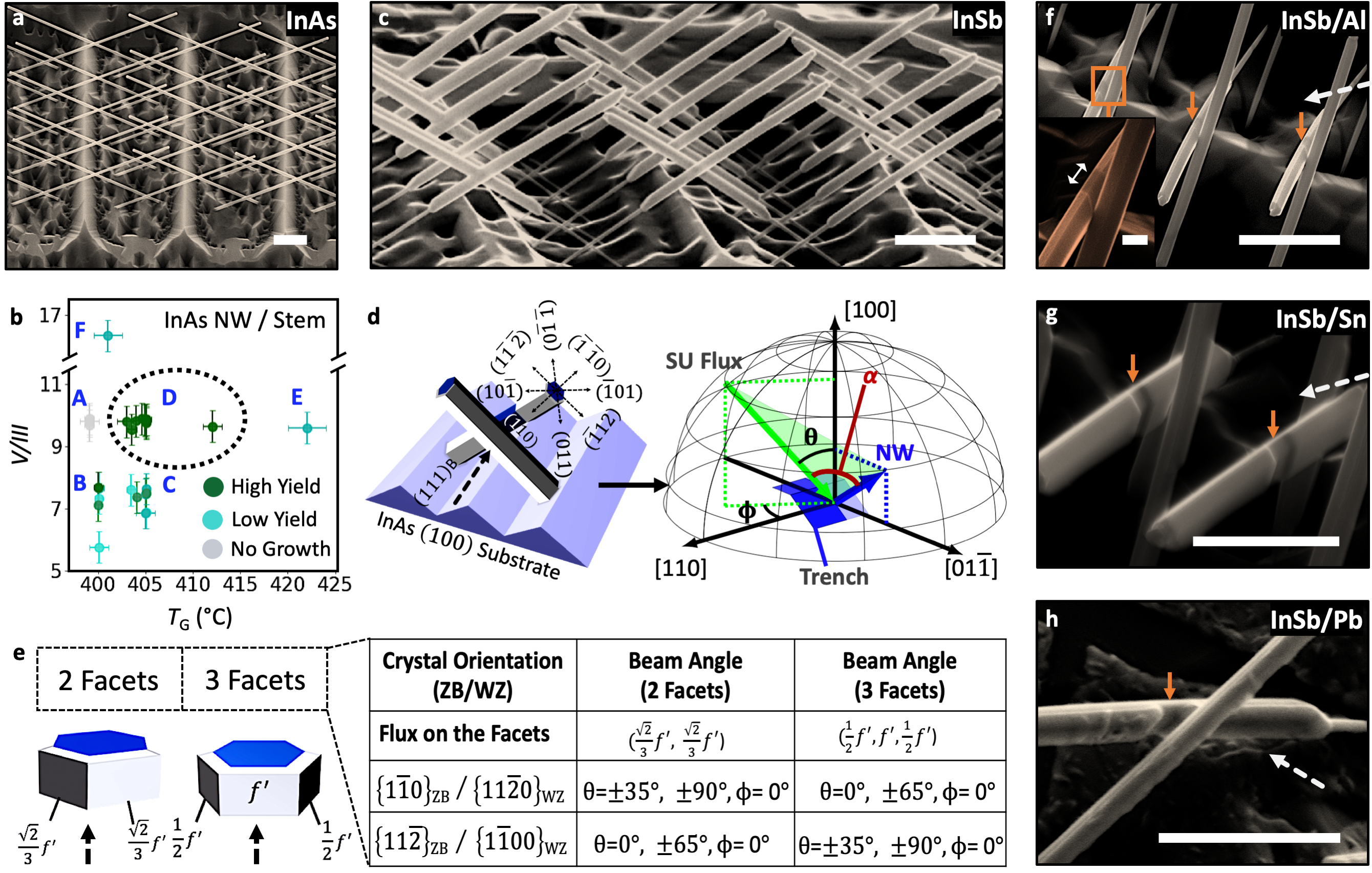}
\vspace{0.2cm}
\caption{\textbf{Semiconductor-superconductor nanowires with shadow junctions.} \textbf{a}, Scanning electron micrograph (SEM, 30$^{\circ}$ tilted) of Au-assisted InAs NW array grown on the pre-processed "V" groove (111)B faceted InAs trenches. \textbf{b}, V/III ratio as a function of InAs NW growth temperature ($T_\mathrm{G}$). The plot is divided into six regions (A-F), and NW growth outcome (yield, morphology, etc.) of each region is investigated. Region "D" ($T_\mathrm{G}$ $\sim$ 401-415$^{\circ}$C with V/III ratio $\sim$ 9-10.5) shows the highest yield and uniform InAs NW growth (dark green circles). \textbf{c}, SEM image of InSb NWs grown from InAs stems. \textbf{d}, Schematic with crystal orientation of the hybrid NWs. Dashed arrow shows the superconductor deposition direction. In the right, schematic of superconductor deposition geometry on the NW with respect to the beam flux direction and the NW growth axis. Here, $\alpha$ is the angle between these two vectors to calculate effective flux ($f'$) on the NW facet. \textbf{e}, Schematic of 3-facets and 2-facets superconductor on the NWs. The table provides the beam angle requirements in the case when $\phi$=0, for 2-facets and 3-facets superconductor coverage. \textbf{f}, Tilted SEM image of InSb NWs with epitaxially grown Al. \textbf{g}, SEM image of InSb/Sn NWs with junctions. \textbf{h}, SEM image of InSb/Pb NW with junction. Orange arrows indicate the shadowed junctions. Dashed arrows indicate the direction of superconductor deposition. Scale bars for (a), (c), (f), (g) and (h) are 1 \textmu m. Scale bar for (f) inset is 100 nm.}
\label{fig1}
\end{figure*}

Josephson junctions (JJs) are critical circuit elements superconducting quantum computing. Gate tunable junctions based on proximitized semiconducting segments in hybrid semiconductor (SE)-superconductor (SU) materials are an interesting class of junctions with potential as JJ elements in transmons qubits \citen{casparis2018superconducting, Larsen2017} as well as critical operators in topological qubits \citen{karzig2017scalable}. Similar to all semiconducting circuit elements, they are highly susceptible to disorder and require dedicated optimization for development towards high fidelity gate operations. In the case of topological quantum computing, achieving disorder free junctions is desirable on several levels. Coupling a one-dimensional semiconductor with strong spin-orbit interaction and large Land\'{e}-g factor to a conventional s-wave superconductor has the fundamental ingredients to generate topologically protected Majorana bound states (MBS) \citen{Lutchyn2010, oreg2010helical, Lutchyn2018, Stanescu2013, Leijnse2012}. If the material fulfills the set of requirements, MBS are expected to tolerate local perturbations and therefore makes it a promising candidate for scalable quantum computing \citen{Nayak2008, karzig2017scalable, aasen2016milestones}. In recent years, there has been significant progress in the development of epitaxially grown SE-SU hybrid materials to fulfill these requirements \citen{krogstrup2015epitaxy, sestoft2017hybrid, KrizekSAG2018, liu2019semiconductor, gill2018selective}. Even though electronic tunnel spectroscopy has yielded results that have been interpreted as signatures of MBS \citen{Mourik2012, Das2012, Rokhinson2012,  deng2012anomalous, Finck2013, Deng2016, Zhang2018, Chen2017, Gul2018}, direct evidence for topologically protected MBS is still missing. Complications in the process of verifying the MBS with tunnel spectroscopy relate not only to the hybrid SE-SU nanowire (NW) segments but also the tunnel junctions which may contain random disorders that may give rise to local Andreev bound states that mimic the zero-bias conductance peaks expected from MBS \citen{PhysRevB.97.214502, lee2014spin, pan2019zerobias}. Avoiding such misinterpretations is certainly desirable and a crucial reason to aim for disorder free junctions. Junctions in spectroscopy devices have been demonstrated with a top-down etching process \citen{Albrecht2016, Deng2016, Nichele2017}, while a more recent alternative approach has been using an \textit{in-situ} shadow method \citen{krizek2017growth, gazibegovic2017epitaxy, carrad2019shadow}. The \textit{in-situ} process leaves the SE surface as it was grown, which seems to be an ideal fabrication approach for gate tunable junctions even though a detailed analysis of the junction formation along with correlations between fabrication and junction quality are still missing.

In this work, we study the synthesis of stacking-fault free InAs, InSb and $\mathrm{InAs_{1-x}Sb_x}$ NWs with epitaxially grown superconductors containing shadowed junctions in single-step growth process using molecular beam epitaxy (MBE). To obtain shadowed junctions at predefined positions we use (111)B faceted trenches on InAs (100) substrates for NWs growth \citen{Dalacu2013, gazibegovic2017epitaxy}. This method provides freedom for controlled positioning of the shadow junctions due to the specified NW growth directions. We study the formation of junctions as a function of the inter-wire distance between the shadowing NW and the junction NW. We also analyze the junction profile, which directly depends on the flux distribution from the source and geometry of the shadowing. Further with different superconductors we investigate the influence of growth kinetics on the junction sharpness. Developing a pre-growth substrate fabrication process including optimized growth and shadow conditions, we demonstrate high junction transparency with reproducible quantized transport. Correlations between the structural and electronic properties of the junctions are done by statistical characterization of the junction morphology and transport properties of junction NWs from selected positions on the growth substrate. We compare \textit{in-situ} shadowed and etched junctions on statistical ensembles of NW devices as well as on the same $\mathrm{InAs_{1-x}Sb_x}$/Al NWs and confirm the superior electrical quality for the shadowed junctions. Finally, measurements at mK temperatures show an $I_\mathrm{C} R_\mathrm{N}$ products over 7 samples close to the KO$-$2 and KO$-$1 limit. Voltage bias measurements reveal the size of the induced superconducting gap, as well as a phase coherence of at least 5 times of the junction length.



The NW substrates are fabricated using electron beam lithography (EBL) and wet-etching process to form (111)B facets in planar InAs (100) substrates, where the Au catalysts are positioned with a subsequent EBL process. As opposed to earlier works \citen{Dalacu2013, gazibegovic2017epitaxy}, we do not use any masks to confine the Au particles, which significantly reduces the pre-processing efforts and minimizes contamination during fabrication. This makes the process suitable for exploring different material combinations with high throughput. Figure \ref{fig1}\textbf{a} shows an SEM image of InAs NWs grown from (111)B trenches. However, the NW growths on the trenches require careful optimization of the growth conditions. We find out that $\mathrm{As_4}$ beam flux is necessary to enhance the initial NW growth rate to escape the competition with planar growth in the trenches. In Fig. \ref{fig1}\textbf{b} we show a map of InAs NW growth yield, which resembles the design of growth parameter optimization. Dark to light green color represents high (> $90\%$) to low yield (< $50\%$) growth of NWs and gray resembles no growth. To help tuning in the right growth parameter space we distinguish between six growth parameter regions which are discussed in supplementary information S2. The growth temperature window in region "D" with V/III ratio $\sim$ 9-10.5 provides the highest yield and uniform morphology of the NWs, while outside of this region the growth either has uneven yield issue or non-uniform NWs. We mostly grow in the lower side of the growth temperature window in region "D", as we get pure wurtzite (WZ) crystal structure at lower temperatures, verified by TEM analysis (see supplementary information S3). Confirming previous reports \citen{Caroff2008}, we are also not able to grow InSb NWs directly from the InAs substrate. It is speculated that the Au-alloy forms a small contact angle to the substrate when Sb is present, which prevents initiation of NW growth \citen{Caroff2008}. However, once the InAs NW stem is formed it is possible to switch into InSb NW growth. With optimized growth conditions (in region "D"), we achieve high yield InSb NWs across the substrate as shown in Fig. \ref{fig1}\textbf{c}. 

Figure \ref{fig1}\textbf{d} shows the schematic of the NWs with shadow junction as grown on the substrate. On the right, the hemisphere diagram shows the coordinates use to describe the superconductor beam flux direction with respect to NWs. Depending on the angle of the incoming flux and orientation of the NW facets, the superconductor can be grown on selected facets. The table in Fig. \ref{fig1}\textbf{e} contains information of beam flux angles required for 2-facet and 3-facet superconductor coverage on $\mathrm{[1\bar{1}0]_{ZB}}$/$\mathrm{[11\bar{2}0]_{WZ}}$ and $\mathrm{[11\bar{2}]_{ZB}}$/$\mathrm{[1\bar{1}00]_{WZ}}$ oriented NWs. The amount of superconductor that is grown on each facet (for a given growth time) is proportional to the effective beam flux ($f'$) on the selected facet. Here $f'$ is defined as the flux impinging on the mid-facet facing the source during 3-facet deposition (see Fig. \ref{fig1}\textbf{e}). If we consider 2-facet depositions then both facets receive equal amounts of material, as an instance, for $\phi$=0 and $\theta$=35$^{\circ}$, $f_{[\bar{1}01]}=f_{[011]}={\frac{\sqrt3}{2}}{f'}$. For 3-facet depositions, the facet facing towards the beam receives $f'$, while the adjacent facets get $f'/2$. An SEM image of InSb/Al NWs with \textit{in-situ} junctions are demonstrated in Fig. \ref{fig1}\textbf{f}, where the inset shows a $\sim$ 100 nm long junction. Because the NW positions are controlled, it is possible to design multiple junctions in a single NW, see supplementary information S5 for different junction schemes. 

Synthesis of hybrid InSb/Sn NWs with junctions has recently been reported \citen{pendharkar2019parity}, showing a route for deposition of alternative superconductors on bare semiconductor NW facets. Due to the simplicity of the single UHV-step process demonstrated in this work, it is easy to not only vary semiconductor composition but also the superconducting materials, providing a versatile platform for exploring wide range of hybrid material combinations. Two promising superconductor alternatives to Al are Sn and Pb, which both have higher $T_c$ (for bulk it is around 3.7 K for Sn and 7.2 K for Pb, compared to 1.2 K for Al). As these superconducting materials are challenging to etch selectively without damaging the semiconducting NW segments, the shadowing method may be critical for the realization of high-quality junctions. In Fig. \ref{fig1}\textbf{g} and Fig. \ref{fig1}\textbf{h} we show Sn and Pb phases grown on InSb and $\mathrm{InAs_{1-x}Sb_x}$ NWs, respectively. After the semiconductor NW growth, the Sn and Pb are grown on liquid nitrogen cooled stage in a UHV chamber connected to the MBE. The shadowing and formation of the junction details will be discussed below.



\begin{figure}[t!]
\vspace{0.2cm}
\includegraphics[scale=0.56]{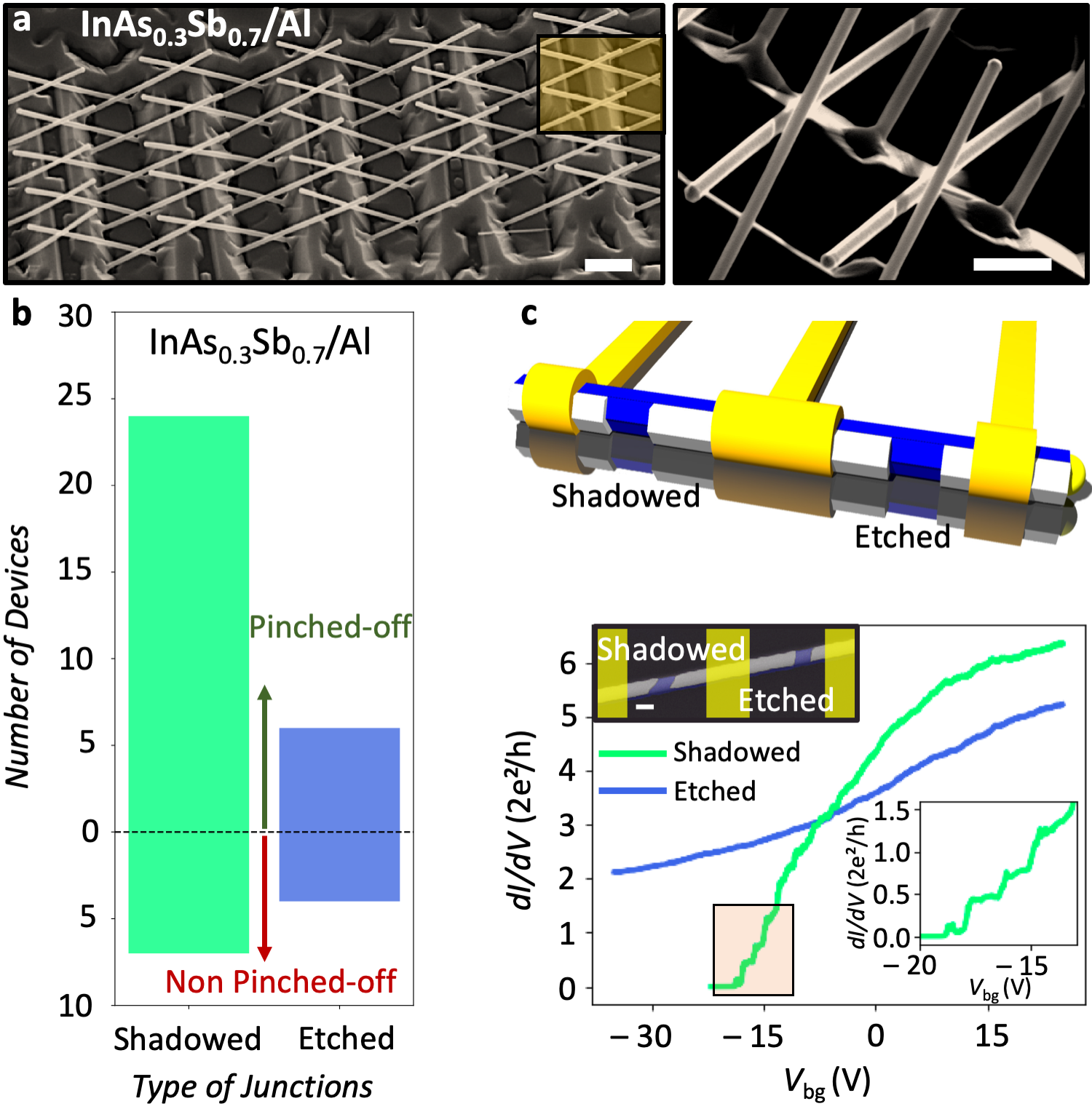}
\vspace{0.2cm}
\caption{\textbf{Gate response statistics of shadowed and etched junctions in InAs$_{0.3}$Sb$_{0.7}$/Al nanowires.} \textbf{a}, SEM (30$^{\circ}$ tilted) of $\mathrm{InAs_{0.3}Sb_{0.7}}$/Al NW array (scale bar is 1\textmu m). The highlighted section shows double shadowed junctions on the NWs (scale bar is 500 nm). \textbf{b}, Pinch-off statistics of the back-gated devices with shadowed (green) and etched (blue) junctions. The y-axis contains the number of devices where the upper side of the dotted line shows devices that are pinched-off and lower side shows devices that are not pinched-off. \textbf{c}, Schematic of the test device with comparable size shadowed and etched junctions in a single NW. Below, electrical measurements of the test device where conductance is shown as a function of gate voltage for shadowed (green) and etched (blue) junctions. The shadowed junction shows quantized plateaus as highlighted with inset. Applied magnetic field, \textbf{B}=6 T in both cases. SEM image of the exact NW is shown inset with a scale bar of 100 nm.} 
\label{fig2}
\end{figure}

Having growth conditions for InAs and InSb NWs on the trenches, the As and Sb fluxes can be tuned to grow $\mathrm{InAs_{1-x}Sb_x}$/Al NWs \citen{sestoft2017hybrid} as shown in Fig. \ref{fig2}\textbf{a}. Similar to InSb NWs, we initiate the $\mathrm{InAs_{1-x}Sb_x}$ NW growth with an InAs stem (using the recipe from \ref{fig1}\textbf{b}, region "D"). The InAs stem is not visible in Fig. \ref{fig2}\textbf{a} due to the over-growth on the substrate. To enhance spin-orbit interaction \citen{winkler2016} while maintaining an efficient field effect response by keeping the carrier density low \citen{sestoft2017hybrid, potts2016twinning} we aim at for Sb composition around x=0.7 (nominal Sb/As flux ratio of 0.8). The composition of the InAs$_{0.3}$Sb$_{0.7}$/Al NWs are measured applying Vegard's law \citen{vegard1921konstitution} for the lattice parameter in ternary alloys (see supplementary information S6 for STEM-EDX analysis). 

These InAs$_{0.3}$Sb$_{0.7}$/Al NWs are used to compare the field-effect response of etched and shadowed junctions as shown in Fig. \ref{fig2}\textbf{b} and \ref{fig2}\textbf{c}. The challenge for etched junctions is to find conditions that selectively etch Al while leaving the semiconductor unharmed. As for instance, we have not been able to find selective etch conditions for Al on InSb. For InAs$_{0.3}$Sb$_{0.7}$ and InAs NWs, we use etch conditions which were optimized in previous studies and apparently leave the semiconductor intact \citen{sestoft2017hybrid, chang2015hard} (see supplementary information S7 for details). Because the electron transport characteristics vary from device to device, we need statistics for comparing quality measures. For this purpose, we compare 41 back-gated devices, 31 with shadowed and 10 with etched junctions, see Fig. \ref{fig2}\textbf{b}. For etched junctions, 7 out of the 10 devices are first measured at 20 mK out of which 6 pinch-off with a threshold voltage of $-3 \pm 1 $ V, a mean saturation conductance of $1.6 \pm 0.2$ $ 2e^2/h$ and the field-effect mobility $\mathrm{\mu_{FE}\approx 1900 \pm 600}$ $\mathrm{cm^2/Vs}$ (highest $\mathrm{\mu_{FE}\approx 4400}$ $\mathrm{cm^2/Vs}$). As a comparison, we measure 9 shadowed devices under identical conditions where 4 devices pinch-off with a mean threshold voltage of $-36.0 \pm 2.5 $ V and rest haven't pinched-off within the applied gate voltage. The mean saturation conductance of all shadowed devices is $9\pm 1$ $2e^2/h$. 

For a higher throughput we turn to measurement at 2 K and measure 22 devices with shadowed junctions where 20 devices pinch-off with a mean $\mathrm{\mu_{FE}\approx 17000 \pm 400}$ $\mathrm{cm^2/Vs}$ (highest $\mathrm{\mu_{FE}\approx 35000}$ $\mathrm{cm^2/Vs}$). The mean threshold voltage and saturation conductance of these devices are $-13 \pm 2 $ V and $5.1 \pm 0.5$ $2e^2/h$ respectively. In comparison, three etched junctions are characterized at 2 K showing a low gate response and no pinch-off within the voltage limits of the system and with mean saturation conductance of $4.1 \pm 0.7$ $2e^2/h$. In short, according to the statistics presented above, shadow junctions exhibit a significantly higher conductance and mobility compared to the etched junctions. 

As an additional comparison we fabricate 11 devices with comparable sized etched and shadowed junctions in the same InAs$_{0.3}$Sb$_{0.7}$/Al NW (shown in Figure \ref{fig2}\textbf{c}). Among them, only three devices are functional on both sides, where the electrical measurement of one of these devices is demonstrated in Fig. \ref{fig2}\textbf{c}. Despite the almost identical appearance of the etched and shadowed junctions in these devices, a radical difference is observed in the transport properties. Here the shadow junction pinched off with clear quantized conductance plateaus around magnetic field, \textbf{B} > 4 T, while the etched junction is not pinched-off within the available voltage range. Similar differences are observed for the other two devices presented in supplementary information S8. We note that there can be many reasons for disorders associated with etching such as undercut during wet etching, increased SE surface roughness, impurities left from etchant and leftovers from the etched metal (see supplementary information S7). There are most likely ways to improve etch recipes, however, unless the etch actively provides protection, it is reasonable to assume that optimized shadowed junctions will generally be of the highest possible quality. Additionally, besides obtaining higher quality junctions, shadowed junctions allow flexibility in the choice of material combinations, where selective etching may be infeasible. Based on this and the above results, we choose to solely focus on shadow junctions for all material combinations.


\begin{figure*}[ht!]
\vspace{0.2cm}
\includegraphics[scale=0.5]{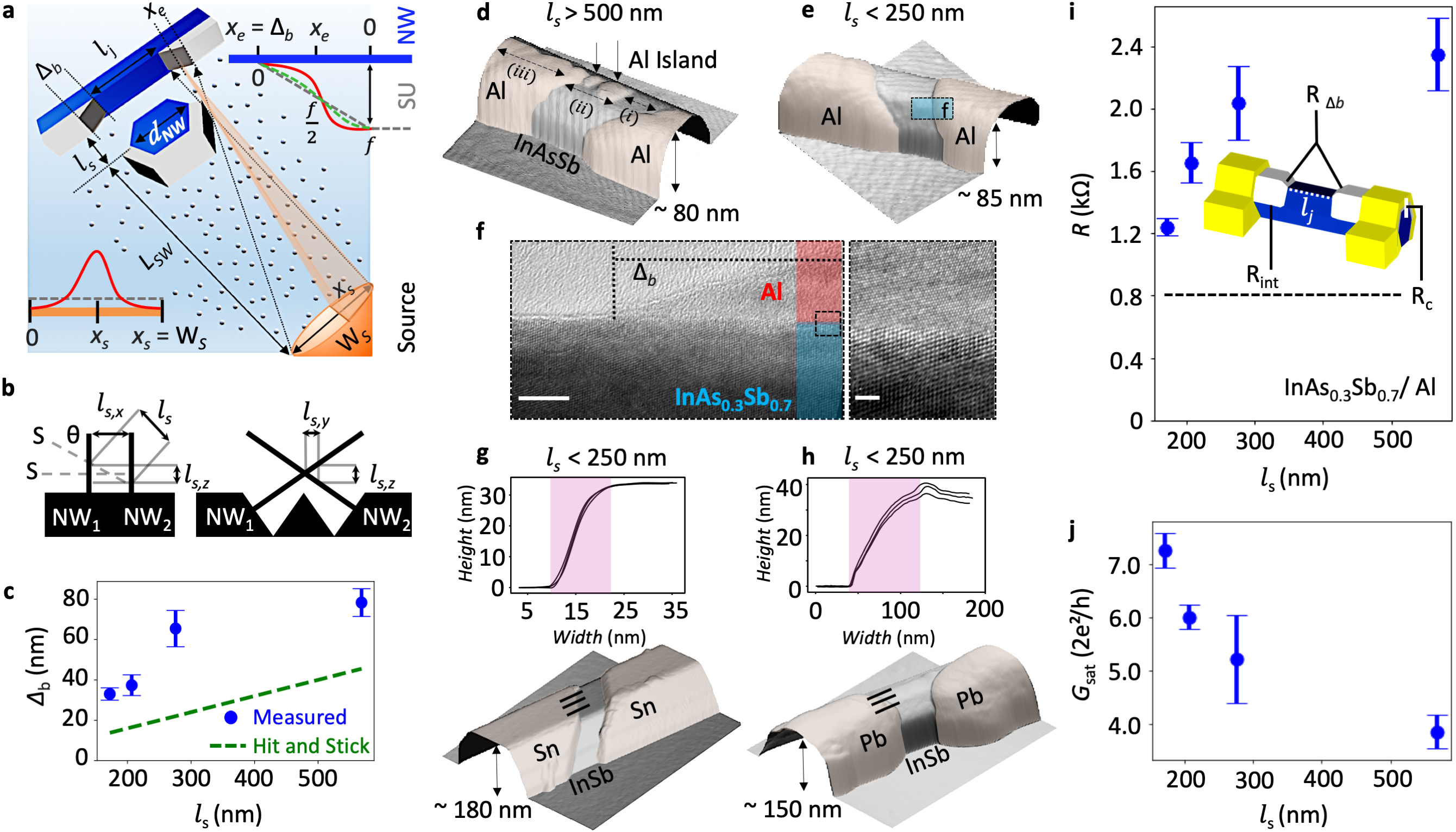}
\vspace{0.2cm}
\caption{\textbf{Edge profile effect on the junction performance.} \textbf{a}, Schematic of the SE-SU junction formation and the edge profile. The broadening ($\Delta_b$) and profile of the junction edge can be determined by the flux distribution in the transition region, inter-wire distances ($l_s$), source to wire distance ($L_{SW}$) and effective width of the source ($W_s$). The junction length ($\ell_j$) depends on the size of the $\Delta_b$ and the diameter of the shadowing NW ($d_{NW}$). (diagram is not drawn to scale). \textbf{b}, Geometry for determining the shadow position. \textbf{c}, $\Delta_b$ as a function of $l_s$. Blue dots are measured $\Delta_b$ for Al junctions on InAs$_{0.3}$Sb$_{0.7}$ NWs and green dashed lines are calculated $\Delta_b$. \textbf{d}, Atomic force micrograph (AFM) of the InAs$_{0.3}$Sb$_{0.7}$/Al junction for $l_s$ > 500 nm. Large Al broadening with multiple Al grains are observed in the junction. Depending on the effective flux distribution on the transition region the junction is divided into three segments. \textbf{e}, AFM of the InAs$_{0.3}$Sb$_{0.7}$/Al NW junction for $l_s$ < 250 nm, where the junction is clean with sharp-edge profile. \textbf{f}, Zoomed-in TEM image from (\textbf{e}) shows the epitaxial SE-SU interface and small Al broadening in the junction. Scale bars are 5 nm and 1 nm respectively. \textbf{g}, AFM of the sharp-edge InSb/Sn junction shadowed by thinner InSb NW for $l_s$ < 250 nm. The line scans, taken at the positions marker by black lines show the broadening of $\sim$ 13 nm. \textbf{h}, AFM of the InSb/Pb junction for $l_s$ < 250 nm. The line scans show the broadening of $\sim$ 75 nm. \textbf{i}, The gate-independent resistance of the shadow junction devices as a function of $l_s$. Inset is the schematic of standard device where $\mathrm{R_{\Delta_b}}$ is the broadening resistance, $\mathrm{R_c}$ is the contact resistance and $\mathrm{R_{int}}$ is the interface resistance. The dotted line till 0.8 k$\Omega$ represents the statistical value of contact resistance obtained by four-probe measurements. \textbf{j}, The conductance saturation as a function of $l_s$.} 
\label{fig3}
\end{figure*} 



The influence of the junction edge morphology on the junction transparency is studied by comparing devices with varying edge sharpness. The edge sharpness is varied by changing inter-wire distance, $l_s$, between the shadowing and the shadowed NW. The reason is sketched in Fig \ref{fig3}\textbf{a}, where the transition region going from a fully shadowed region to a fully unblocked region (with nominal beam-flux $f$) is given by $\Delta_b=\dfrac{{W_S}l_s}{L_{SW}}$. Here $L_{SW}$ is the "source to wire" distance and $W_S$ is the width of the source opening. The effective flux in the transition region $f'(x_e)$ as a function of coordinate $0<x_e<\Delta_b$ is directly related to the flux distribution across the source opening as $f'(x_e)\propto \int_0^{W_S-\frac{L_{SW}}{l_s}x_e}f_{source}dr$. Here $f_{source}$ is the beam-flux originating from a point source within a cutoff area up to $W_S-\frac{L_{SW}}{l_s}x_e$ of the source opening (cutoff determined by $x_e$) as shown in Fig \ref{fig3}\textbf{a}. Thus, a point in the transition region $x_e$ sees a fraction of the source from where the effective impinging flux originates. If the outgoing flux distribution within the source opening is uniform, it can be shown that the effective flux in the transition region is given by: $f'(x_e) = \frac{f}{\pi} \large[\frac{\pi}{2} - X_e $   $\sqrt{1-X_e^2}- \arcsin{X_e}\large]$ where, $X_e = \frac{2L_{SW}}{l_s W_S}x_e-1$. Such an outgoing flux distribution originating from an uniform circular source opening would give a flux distribution in the transition region as shown in the dashed green line in the inset of Fig \ref{fig3}\textbf{a}, while an uniform beam estimated from a 2D model (or from a hypothetical squared source opening) would provide a linearly increasing flux distribution as a reasonable approximation (as shown with the dashed black line). On the other hand, for a circular source opening with a Gaussian flux distribution the transition region will see an effective flux closer to a step-profile flux distribution shown as red solid line. 

If the temperature is sufficiently low, such that the adatoms are kinetically limited to stick where they land ("hit and stick" model), the shape of the junction edges will directly map the flux distribution from the source opening as described above. However, the "hit and stick" model collapse if the adatoms are mobile. This will alter the broadening towards equilibrium shaped morphologies. The length scale at which kinetics plays a role can be described by a characteristic adatom migration length \citen{krogstrup2015epitaxy, krogstrup2013advances}: $\lambda_a\propto\sqrt{\frac{1}{\rho_a}}{\exp(-\dfrac{\delta h_{aa}-\delta\mu_{inc}}{k_BT})}$, where, $\rho_a$ is the adatom density, $\delta h_{aa}$ and $\delta\mu_{inc}$ are a characteristic activation barrier for migration and chemical potential of the adatoms respectively. As seen from this equation, also the beam flux ($f$) can play a role on $\lambda_a$, which complicates the analysis of the adatom kinetics due to the effective flux gradient in the transition region. To limit the adatom mobility we grow the SU thin film at low substrate temperatures where $\lambda_a$ is sufficiently short to allow for the formation of a uniform thin film at the given flux.

The determination scheme of shadow location is sketched in Fig. \ref{fig3}\textbf{b} where $l_s$ (and therefore $\Delta_b$) is controlled with two parameters $\theta$ and $l_{s,x}$, the spacing between the Au dots along opposite trenches. For $\phi=0$ the relation is simply $l_s=\frac{l_{s,x}}{\sin{\theta}}$. These two parameters also determine the position of the shadow on the NW via the equations $l_{s,y} = \frac{l_s \cos{\theta}}{\tan{35.3^{\circ}}}$ and $l_{s,z} =  l_s \cos{\theta}$. As described above, $\Delta_b$ depends on $l_{s,x}$ which is controlled by the Au droplet positioning during substrate preparation. If the Au droplets offset on the opposite facets are within $d_{NW}$ range then the NWs will merge to form nano-crosses or other type of networks \citen{krizek2017growth, Dalacu2013}. However, for obtaining sharp-edged junctions $l_s$ needs to be as small as possible without merging. We vary $l_{s}$ from $\sim$ 170 nm to $\sim$ 570 nm from trench to trench on a given substrate and measure the broadening on selected NWs with AFM. Figure \ref{fig3}\textbf{c} shows measured broadening (blue points) together with the calculated "hit and stick" broadening (green dashed line) for InAs$_{0.3}$Sb$_{0.7}$/Al junctions. Here, $W_s$ $\sim$ 1.6 cm and $L_{SW}$ $\sim$ 20 cm for Al deposition in our MBE. The measured mean broadening follows the trend of "hit and stick" model with an offset, which indicates that adatom kinetics plays an important role for Al shadow junction formation under these conditions.

\begin{figure}[t!]
\vspace{0.2cm}
\includegraphics[scale=0.40]{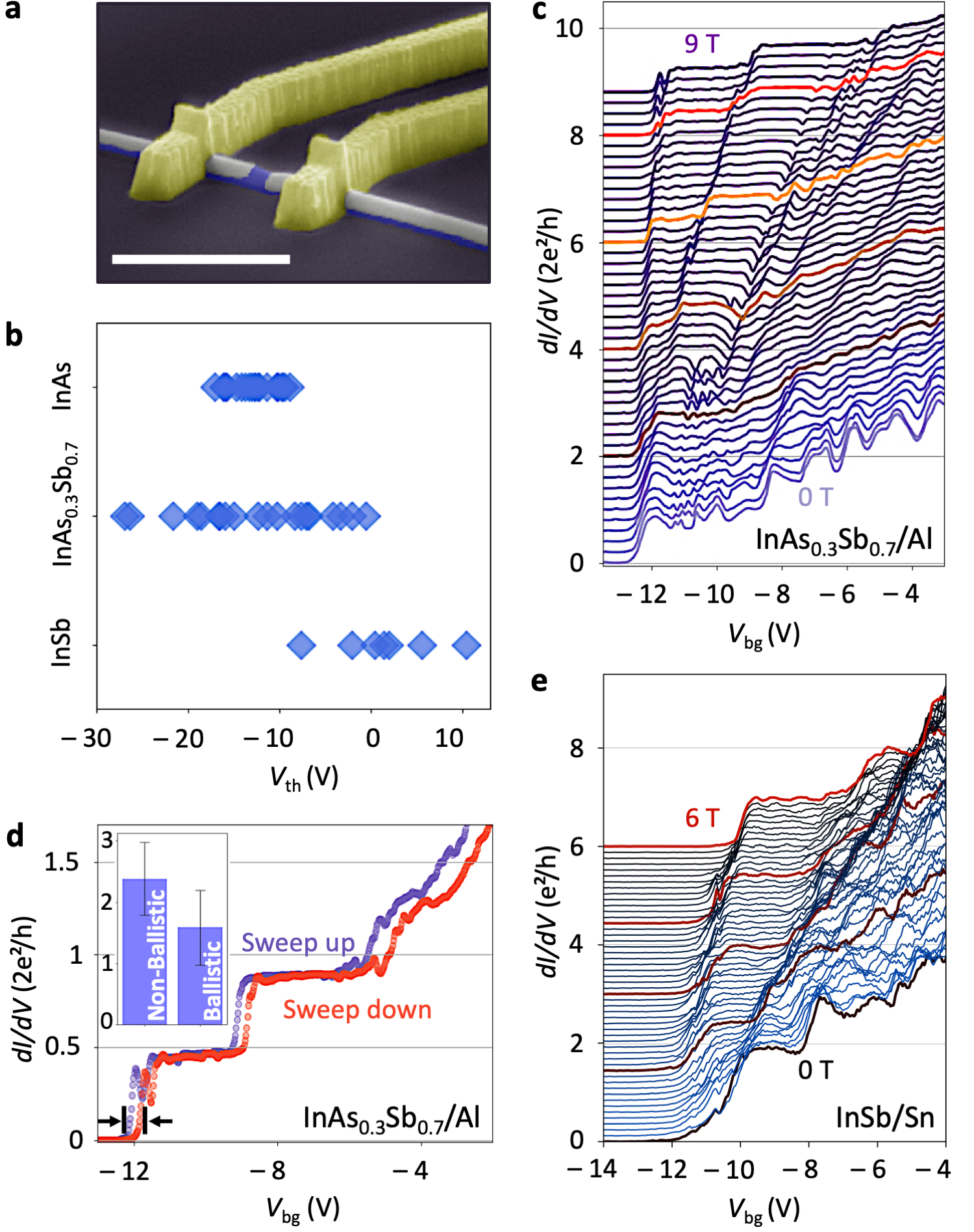}
\vspace{0.2cm}
\caption{\textbf{Quantized transport in sharp-edged junctions.} \textbf{a}, Pseudo colored SEM image of a typical single-junction back-gated device. Scale bar is 1\textmu m. \textbf{b}, Pinch-off voltage statistics for InAs, InAs$_{0.3}$Sb$_{0.7}$ and InSb NW junction devices. \textbf{c}, Differential conductance as a function of gate voltage and magnetic field of a InAs$_{0.3}$Sb$_{0.7}$/Al NW junction. \textbf{d}, Hysteresis of the device shown in (\textbf{c}) where sweep up and down follows closely. Inset is a comparison of a statistical value of hysteresis between quantized and non-quantized devices. \textbf{e}, Differential conductance as a function of gate voltage and magnetic field for InSb/Sn NW junction.}

\label{fig4}
\end{figure} 

A general trend is that small Al islands are formed in the junction region for $l_s >500$ nm (as shown in Fig. \ref{fig3}\textbf{d}). The junction edge broadening with discrete Al islands are estimated from fitting a curve over the measure islands. For $l_s <250$ nm, we observe well defined single junctions with no Al islands as shown in Fig. \ref{fig3}\textbf{e} and confirmed by a TEM image in Fig. \ref{fig3}\textbf{f}. We attribute the larger broadening profile of Al shadow edges, than predicted by the hit-and-stick model, to the kinetically driven equilibrium shape. For $l_s=250$ nm, the Sn and Pb based junctions on InSb exhibit sharp edge shadows as shown in Fig. \ref{fig3}\textbf{g} and Fig. \ref{fig3}\textbf{h}. In contrast to the Al deposition, we used e-beam evaporation of Sn and Pb where the source opening $W_S$ depend on how the electron beam are focused on the targeted materials. In the case of Sn, the effective area were visibly smaller than the area of the total target, which means that the effective source opening $W_S$ and therefore $\Delta_b$ will be smaller for a given $l_s$ in case of "hit and stick" conditions. The outgoing flux distribution can be estimated with a Gaussian profile as discussed in \ref{fig3}\textbf{a} leading to a sharp edge flux profile in the transition region. Fig. \ref{fig3}\textbf{g} with line scale showing $\Delta_b$ $\sim$ 13 nm confirms a sharp edge profile of Sn edge. Here, the measured sharpness may be underestimated due to the AFM tip diameter. On the other hand, for Pb based shadowing, the outgoing flux distribution is more uniform from the source, as a result $\Delta_b$ for Pb based junction is larger than that of Sn, $\sim$ 75 nm as extracted from line cuts in Fig. \ref{fig3}\textbf{h}.

We study correlations between the junction transparency and the critical parameter for the junction profile $l_s$ on the Al shadowed NWs. For this purpose we calculate the gate independent resistance $R$ by fitting the pinch-off curves as described in supplementary information S9. This gate independent resistance contains mainly three contributions: contact resistance ($\mathrm{R_c}$), broadening resistance ($\mathrm{R_{\Delta_b}}$) and SE-SU interface resistance ($\mathrm{R_{int}}$). In Fig. \ref{fig3}\textbf{i}, we can see that the InAs$_{0.3}$Sb$_{0.7}$/Al shadow junction resistance statistically increases with increasing $l_s$. It is surprising that the junction resistance, $\mathrm{R_{\Delta_b}}$ depends on the slope of the Al towards the junction, also for junctions without visible Al islands, however the trend seems significant. For junctions with dewetted Al islands it seem reasonable with a reduced junction transparency due to potential variations caused by Al islands across the junctions. Using standard four-probe measurements, the measured mean junction resistance is $\mathrm{R_c}=0.8$ k$\Omega$ as shown in Fig \ref{fig3}\textbf{i}. Figure \ref{fig3}\textbf{j} shows conductance saturation of the devices decrease with increased $l_s$. We attribute this effect to the junctions with multiple Al grains for $l_s$ > 500 nm. Unexpectedly, we also observe a trend of decreasing conductance for $l_s < 250$ nm, although Al grains do not form within this range. We presume the profile of broadening within that range may play a role on the conductance deviation.


We further investigate the device performances of the sharp edged junctions for InAs, InSb and InAs$_{0.3}$Sb$_{0.7}$ NWs. Pseudo colored SEM image of a typical single shadow junction device is shown in Fig \ref{fig4}\textbf{a}. In Fig \ref{fig4}\textbf{b}, we show the pinch-off voltages for Al based junction devices measured at 2 K. The pinch-off voltages for InAs$_{0.3}$Sb$_{0.7}$/Al junctions show the widest span from $\sim  - 30$V to $\sim$ 0V, while the InAs/Al junctions pinch-off in the range of $\sim$ 0 to $-10$V. On the other hand, InSb/Al junction devices show pinch-off at mainly positive $V_g$. We ascribe the statistical differences to the band alignment between the semiconductor and the Al \citen{schuwalow2019band}. An example of quantized conductance in a InAs$_{0.3}$Sb$_{0.7}$/Al junction device is shown in Fig. \ref{fig4}\textbf{c}, where the conductance is measured as a function of $V_g$ and magnetic field (\textbf{B}). A general trend is that the conductance plateaus are less pronounced at low \textbf{B} but gets gradually sharper with increasing field. This can be ascribed to lower electron back-scattering rates at higher \textbf{B}. The first subband splits into two spin-split subbands due to the Zeeman effect which leads to an energy difference $g\mu_B\textbf{B}$, where $g$ is the Land\'{e}-g factor and $\mu_B$ is Bohr's magneton. For this particular InAs$_{0.3}$Sb$_{0.7}$/Al device we see the emergence of the spin-split sub-bands around \textbf{B} > 2.2 T. However, we generally see the visible splitting appearing around \textbf{B}= 2-3 T. We speculate that the late emergence of visible spin-split bands are related to electron-electron interaction within the two subbands. The obtained quantized values in Fig. \ref{fig4}\textbf{c} are lower than the predicted $Ne^2/h$ because of a constant contact resistance leading to plateaus of $0.45 \cdot 2e^2/h$ (the second at $0.9\cdot 2e^2/h$ and the third one is barely observed at $1.3 \pm 0.5 \cdot 2e^2/h$), suggesting a non changing contact resistance with gate due to the equal quantized values for each subband. Figure \ref{fig4}\textbf{d} demonstrates a hysteresis which is much smaller than subband spacing in the device presented in Fig. \ref{fig4}\textbf{c}. The inset show a comparison of hysteresis statistics near the pinch-off region between quantized and non-quantized devices, where the quantized devices typically exhibit a slightly smaller hysteresis $\sim$ 1.5 V, compare to others ($\sim$ 2.3 V). In Fig. \ref{fig4}\textbf{e} we examine sharp-edge InSb/Sn junction device, with each trace offset by the value of the \textbf{B}. In contrast to the InSb/Al junctions, the devices with Sn show a negative pinch-off voltage around $-10$ V, caused by the different band alignment of Sn to InSb. In these devices, after the subtraction of the filter resistances in the fridge and a constant contact resistance, a clear plateau at $2e^2/h$ is visible even at zero field, suggesting a scattering length on the order of a few hundred nanometers. Furthermore, unlike the sample shown in Fig. \ref{fig4}\textbf{c}, splitting of the subband is visible immediately as the field is increased. A crossing of the first two spin-split subbands is visible at around 2 T, characterized by the disappearance and re-emergence of a plateau at $2e^2/h$. This effect is expected due to the large Land\'{e}-g factor of InSb \citen{van2013quantized}.

\begin{figure}[t!]
\vspace{0.2cm}
\includegraphics[scale=0.47]{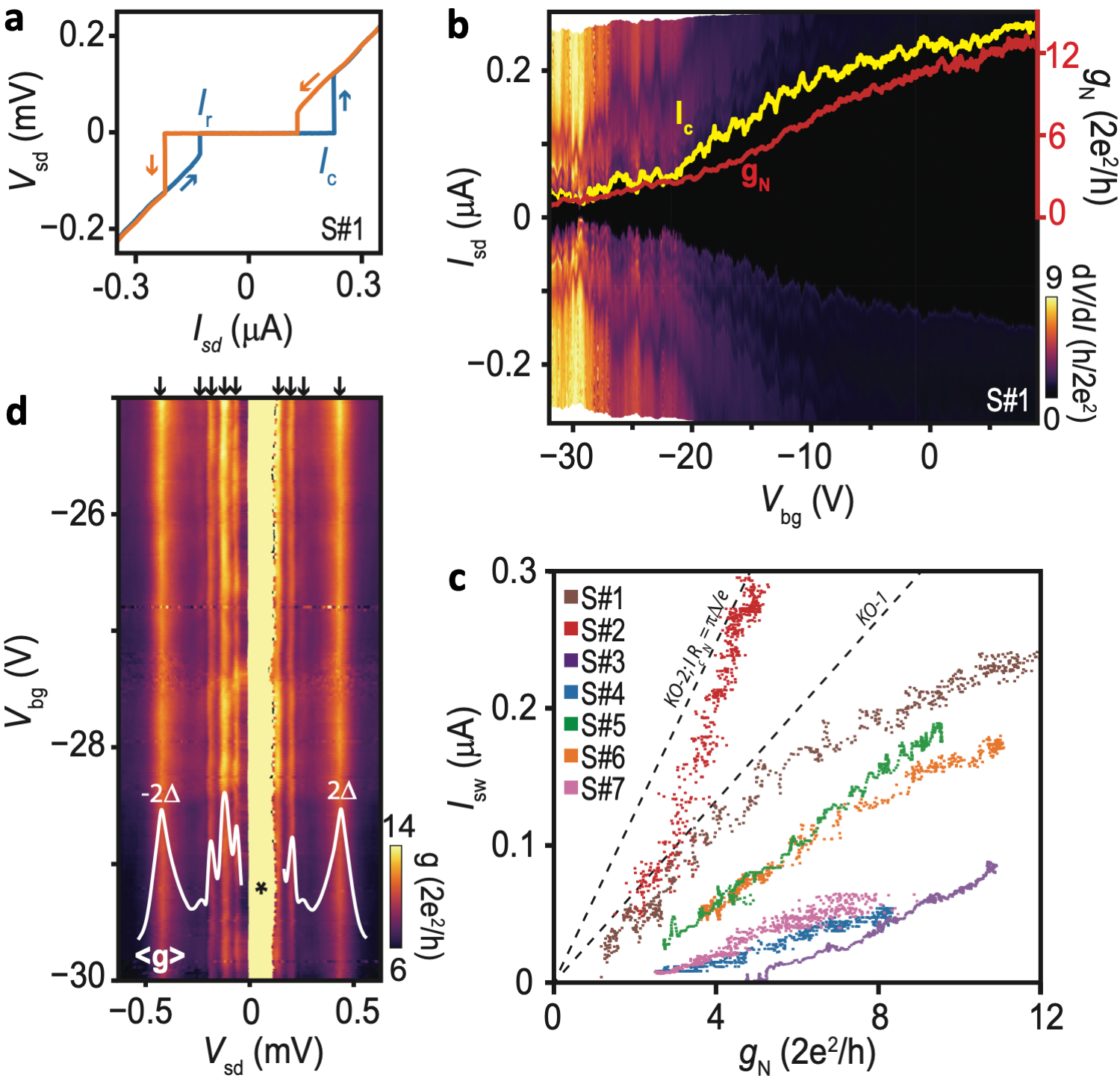}
\vspace{0.2cm}
\caption{\textbf{Supercurrent and multiple Andreev reflections in the sharp-edge junctions.} \textbf{a}, Typical $VI$-curve for an InAs$_{0.3}$Sb$_{0.7}$/Al shadow Josephson junction (Device S$\#$1). The switching and retrapping currents are indicated. \textbf{b}, Differential resistance as a function of current and gate potential. The black region corresponds to the zero-voltage state. $I_\mathrm{C}$ and normal state conductance $g_\mathrm N$ measured at $B_\bot=0.3\,\mathrm T$ are shown. \textbf{c}, $I_\mathrm C$ vs. $g_\mathrm N$ for all measured devices. Dashed lines are the theoretical expectations (see text). \textbf{d}, Voltage-biased measurement of the sub-gap structure of S$\#$1 showing resonances of multiple Andreev reflections. The high-conductance region asymmetric around zero bias ($*$) is related to a supercurrent branch enabled by the finite resistance of the cryostat wiring (see text).}

\label{fig5}
\end{figure} 


Figure \ref{fig5} presents low temperature ($T \sim 20 \, \mathrm{mK}$) electrical measurements performed on seven InAs$_{0.3}$Sb$_{0.7}$/Al shadow JJs (S$\#1$--S$\#$7) with $l_j \sim 100\,\mathrm{nm}$. A single contact to the Al shell is fabricated from Ti/Au normal metal on either side of the shadow junction and split into separate bond-pads to allow for a pseudo four-terminal configuration eliminating contributions from highly resistive filters in the cryostat. The transmission of the junction is tunable by the back-gate potential $V_\mathrm{g}$ and Fig. \ref{fig5}\textbf{a} shows typical $VI$-curves at $V_\mathrm{g} = 0 \, \mathrm{V}$ for sample S$\#$1. A zero-voltage state is observed corresponding to a switching current $I_\mathrm{C}$ exceeding $200\, \mathrm{nA}$ with pronounced hysteresis between up/down sweep directions which is commonly observed in such devices and attributed to heating effects or underdamped junction \citen{doh2005tunable, courtois2008origin}. Figure \ref{fig5}\textbf{b} shows the differential resistance as a function of $I$ and $V_\mathrm{g}$. The zero-resistance state is clearly observed and $I_\mathrm{C}$ decreases with $V_\mathrm{g}$ as the transparency of the $n$-type semiconductor weak link decreases towards pinch-off at $V_\mathrm{g} \sim -40 \, \mathrm V$. Also shown are the extracted gate-dependence of the switching current $I_\mathrm{C}(V_\mathrm{g})$ and the normal state conductance $g_\mathrm{N}(V_\mathrm{g})$ measured at with a magnetic field $B_\bot =0.3 \, \mathrm{T}$ applied perpendicular so the substrate and exceeding the critical field of the superconducting leads. The product of $I_\mathrm{C}$ and $R_\mathrm{N}=1/g_\mathrm{N}$ is a typical voltage characterizing JJs and Fig. \ref{fig5}\textbf{c} shows $I_\mathrm{C}$ vs.\ $g_\mathrm{N}$ for all devices where the range of $g_\mathrm{N}$ is spanned by sweeping $V_\mathrm{g}$. For samples S$\#$2-7 the curves are extracted from the data included in supplementary information S12. The dashed line labeled KO$-$1 (KO$-$2) shows $I_\mathrm{C} R_\mathrm{N} = \pi \Delta/2e$ ($I_\mathrm{C} R_\mathrm{N} = \pi \Delta/e$) expected for a JJ in the short, quasi-ballistic and dirty (ballistic) regime with the mean-free path $l_e \ll l_j \ll \xi$ ($l_e \gtrsim l_j, l_j \ll \xi$), and a superconducting gap $\Delta = 200 \, \mu \mathrm{eV}$ expected for Al and matching voltage-biased measurements discussed below. $\xi$ is the superconducting coherence length \citen{kulik1977properties}. JJs with semiconductor NWs weak links have been the subject of a large number of investigations since the original work of Doh \citen{doh2005tunable} and the critical currents in these devices are generally much lower than the KO$-$1 and KO$-$2 predictions and $I_\mathrm{c} R_\mathrm{n}$ significantly underestimates $\Delta$ \citen{xiang2006ge, nilsson2012supercurrent, gharavi2017nb}. The origin of this suppression is unknown, but has been speculated to arise due to disorder and in-homogeneity or to heavily underdamped junctions. For the InAs$_{0.3}$Sb$_{0.7}$ shadow junctions studied here the critical currents are relatively high, and samples S$\#$1 and S$\#$2 follow approximately the KO$-$1 and the ballistic KO$-$2 result. The remaining devices have suppressed $I_\mathrm{C}$ for high $R_\mathrm N$ indicating presence of channels with weak contribution to the supercurrent. At lower resistance the increase in $I_\mathrm C$ with $g_\mathrm N$ follows the KO$-$1 slope consistent with additional channels with contribution to $I_\mathrm C$ as predicted by the model. We attribute these results to the high quality of the sharp edge InAs$_{0.3}$Sb$_{0.7}$/Al shadow junctions and clean interface. For sample S$\#$1 the phase-coherence is confirmed by the voltage-biased measurement in Fig. \ref{fig5}\textbf{d} which shows a clear $V_\mathrm{g}$-independent sub-gap structure which we attribute to multiple andreev reflections (MAR) as previously studied in NW JJ \citen{doh2005tunable}. The resonance resolved at lowest $V_{sd}$ corresponds approximately to the $n=5^\mathrm{th}$ order MAR $2\Delta/ne$ process requiring five coherent andreev reflection processes. Higher order MAR processes may be present but are inaccessible in these measurements due to the cryostat line resistances $\sim$ 6 k$\Omega$ making the measurement an effective current-biased measurement at low applied voltages.



To conclude, we present a versatile single-step UHV crystal growth method to fabricate epitaxial SE-SU NWs with high quality gate-tunable superconducting junctions. The flexibility of the approach is exemplified with the growth of InAs, InSb and InAs$_{0.3}$Sb$_{0.7}$ NWs with \textit{in-situ} shadowed junctions in Al, Sn and Pb. Based on the performance statistics of field effect InAs$_{0.3}$Sb$_{0.7}$/Al devices we show that the quality of shadowed junctions are significantly higher than the etched junctions. Furthermore, for the shadowed junctions we demonstrate that the junction transparency depends on the junction edge profile. We conclude that the junctions with sharp edges has high transparency, exhibiting extremely large supercurrents and easily resolved quantized conductance of the lowest subbands. This study shows a path towards reliable gate-tunable operations in superconducting quantum networks.




\section{Acknowledgement}
The project is supported by European Union Horizon 2020 research and innovation program under the Marie Sk\l{}odowska-Curie Grant No. 722176 (INDEED), Microsoft Quantum and the European Research Council (ERC) under Grant No. 716655 (HEMs-DAM). Authors acknowledge C. B. S\o{}rensen for maintenance and help with the MBE system. Thanks to Shivendra Upadhyay, Robert McNeil technical support in NBI cleanroom. Also, thanks to Philippe Caroff, Keita Otani, Toma\v{s} Stankevi\v{c}, Emrah Yucelen for helpful discussions during this research.


\section{Competing financial interests}
The authors declare no competing financial interests.

\section{Supplementary Information}

The Supplementary Information is available at: \url{https://sid.erda.dk/share_redirect/FAS4l1InS3}

\bibliography{ref}

\begin{thebibliography}{48}%
\makeatletter
\providecommand \@ifxundefined [1]{%
 \@ifx{#1\undefined}
}%
\providecommand \@ifnum [1]{%
 \ifnum #1\expandafter \@firstoftwo
 \else \expandafter \@secondoftwo
 \fi
}%
\providecommand \@ifx [1]{%
 \ifx #1\expandafter \@firstoftwo
 \else \expandafter \@secondoftwo
 \fi
}%
\providecommand \natexlab [1]{#1}%
\providecommand \enquote  [1]{``#1''}%
\providecommand \bibnamefont  [1]{#1}%
\providecommand \bibfnamefont [1]{#1}%
\providecommand \citenamefont [1]{#1}%
\providecommand \href@noop [0]{\@secondoftwo}%
\providecommand \href [0]{\begingroup \@sanitize@url \@href}%
\providecommand \@href[1]{\@@startlink{#1}\@@href}%
\providecommand \@@href[1]{\endgroup#1\@@endlink}%
\providecommand \@sanitize@url [0]{\catcode `\\12\catcode `\$12\catcode
  `\&12\catcode `\#12\catcode `\^12\catcode `\_12\catcode `\%12\relax}%
\providecommand \@@startlink[1]{}%
\providecommand \@@endlink[0]{}%
\providecommand \url  [0]{\begingroup\@sanitize@url \@url }%
\providecommand \@url [1]{\endgroup\@href {#1}{\urlprefix }}%
\providecommand \urlprefix  [0]{URL }%
\providecommand \Eprint [0]{\href }%
\providecommand \doibase [0]{http://dx.doi.org/}%
\providecommand \selectlanguage [0]{\@gobble}%
\providecommand \bibinfo  [0]{\@secondoftwo}%
\providecommand \bibfield  [0]{\@secondoftwo}%
\providecommand \translation [1]{[#1]}%
\providecommand \BibitemOpen [0]{}%
\providecommand \bibitemStop [0]{}%
\providecommand \bibitemNoStop [0]{.\EOS\space}%
\providecommand \EOS [0]{\spacefactor3000\relax}%
\providecommand \BibitemShut  [1]{\csname bibitem#1\endcsname}%
\let\auto@bib@innerbib\@empty
\bibitem [{\citenamefont {Casparis}\ \emph {et~al.}(2018)\citenamefont
  {Casparis}, \citenamefont {Connolly}, \citenamefont {Kjaergaard},
  \citenamefont {Pearson}, \citenamefont {Kringh{\o}j}, \citenamefont {Larsen},
  \citenamefont {Kuemmeth}, \citenamefont {Wang}, \citenamefont {Thomas},
  \citenamefont {Gronin} \emph {et~al.}}]{casparis2018superconducting}%
  \BibitemOpen
  \bibfield  {author} {\bibinfo {author} {\bibfnamefont {L.}~\bibnamefont
  {Casparis}}, \bibinfo {author} {\bibfnamefont {M.~R.}\ \bibnamefont
  {Connolly}}, \bibinfo {author} {\bibfnamefont {M.}~\bibnamefont
  {Kjaergaard}}, \bibinfo {author} {\bibfnamefont {N.~J.}\ \bibnamefont
  {Pearson}}, \bibinfo {author} {\bibfnamefont {A.}~\bibnamefont
  {Kringh{\o}j}}, \bibinfo {author} {\bibfnamefont {T.~W.}\ \bibnamefont
  {Larsen}}, \bibinfo {author} {\bibfnamefont {F.}~\bibnamefont {Kuemmeth}},
  \bibinfo {author} {\bibfnamefont {T.}~\bibnamefont {Wang}}, \bibinfo {author}
  {\bibfnamefont {C.}~\bibnamefont {Thomas}}, \bibinfo {author} {\bibfnamefont
  {S.}~\bibnamefont {Gronin}},  \emph {et~al.},\ }\href@noop {} {\bibfield
  {journal} {\bibinfo  {journal} {Nature Nanotechnology}\ }\textbf {\bibinfo
  {volume} {13}},\ \bibinfo {pages} {915} (\bibinfo {year} {2018})}\BibitemShut
  {NoStop}%
\bibitem [{\citenamefont {Larsen}\ \emph {et~al.}(2015)\citenamefont {Larsen},
  \citenamefont {Petersson}, \citenamefont {Kuemmeth}, \citenamefont
  {Jespersen}, \citenamefont {Krogstrup}, \citenamefont {Nyg\aa{}rd},\ and\
  \citenamefont {Marcus}}]{Larsen2017}%
  \BibitemOpen
  \bibfield  {author} {\bibinfo {author} {\bibfnamefont {T.~W.}\ \bibnamefont
  {Larsen}}, \bibinfo {author} {\bibfnamefont {K.~D.}\ \bibnamefont
  {Petersson}}, \bibinfo {author} {\bibfnamefont {F.}~\bibnamefont {Kuemmeth}},
  \bibinfo {author} {\bibfnamefont {T.~S.}\ \bibnamefont {Jespersen}}, \bibinfo
  {author} {\bibfnamefont {P.}~\bibnamefont {Krogstrup}}, \bibinfo {author}
  {\bibfnamefont {J.}~\bibnamefont {Nyg\aa{}rd}}, \ and\ \bibinfo {author}
  {\bibfnamefont {C.~M.}\ \bibnamefont {Marcus}},\ }\href@noop {} {\bibfield
  {journal} {\bibinfo  {journal} {Physical Review Letters}\ }\textbf {\bibinfo
  {volume} {115}},\ \bibinfo {pages} {127001} (\bibinfo {year}
  {2015})}\BibitemShut {NoStop}%
\bibitem [{\citenamefont {Karzig}\ \emph {et~al.}(2017)\citenamefont {Karzig},
  \citenamefont {Knapp}, \citenamefont {Lutchyn}, \citenamefont {Bonderson},
  \citenamefont {Hastings}, \citenamefont {Nayak}, \citenamefont {Alicea},
  \citenamefont {Flensberg}, \citenamefont {Plugge}, \citenamefont {Oreg} \emph
  {et~al.}}]{karzig2017scalable}%
  \BibitemOpen
  \bibfield  {author} {\bibinfo {author} {\bibfnamefont {T.}~\bibnamefont
  {Karzig}}, \bibinfo {author} {\bibfnamefont {C.}~\bibnamefont {Knapp}},
  \bibinfo {author} {\bibfnamefont {R.~M.}\ \bibnamefont {Lutchyn}}, \bibinfo
  {author} {\bibfnamefont {P.}~\bibnamefont {Bonderson}}, \bibinfo {author}
  {\bibfnamefont {M.~B.}\ \bibnamefont {Hastings}}, \bibinfo {author}
  {\bibfnamefont {C.}~\bibnamefont {Nayak}}, \bibinfo {author} {\bibfnamefont
  {J.}~\bibnamefont {Alicea}}, \bibinfo {author} {\bibfnamefont
  {K.}~\bibnamefont {Flensberg}}, \bibinfo {author} {\bibfnamefont
  {S.}~\bibnamefont {Plugge}}, \bibinfo {author} {\bibfnamefont
  {Y.}~\bibnamefont {Oreg}},  \emph {et~al.},\ }\href@noop {} {\bibfield
  {journal} {\bibinfo  {journal} {Physical Review B}\ }\textbf {\bibinfo
  {volume} {95}},\ \bibinfo {pages} {235305} (\bibinfo {year}
  {2017})}\BibitemShut {NoStop}%
\bibitem [{\citenamefont {Lutchyn}\ \emph {et~al.}(2010)\citenamefont
  {Lutchyn}, \citenamefont {Sau},\ and\ \citenamefont {Sarma}}]{Lutchyn2010}%
  \BibitemOpen
  \bibfield  {author} {\bibinfo {author} {\bibfnamefont {R.~M.}\ \bibnamefont
  {Lutchyn}}, \bibinfo {author} {\bibfnamefont {J.~D.}\ \bibnamefont {Sau}}, \
  and\ \bibinfo {author} {\bibfnamefont {S.~D.}\ \bibnamefont {Sarma}},\
  }\href@noop {} {\bibfield  {journal} {\bibinfo  {journal} {Physical Review
  Letters}\ }\textbf {\bibinfo {volume} {105}},\ \bibinfo {pages} {077001}
  (\bibinfo {year} {2010})}\BibitemShut {NoStop}%
\bibitem [{\citenamefont {Oreg}\ \emph {et~al.}(2010)\citenamefont {Oreg},
  \citenamefont {Refael},\ and\ \citenamefont {von Oppen}}]{oreg2010helical}%
  \BibitemOpen
  \bibfield  {author} {\bibinfo {author} {\bibfnamefont {Y.}~\bibnamefont
  {Oreg}}, \bibinfo {author} {\bibfnamefont {G.}~\bibnamefont {Refael}}, \ and\
  \bibinfo {author} {\bibfnamefont {F.}~\bibnamefont {von Oppen}},\ }\href@noop
  {} {\bibfield  {journal} {\bibinfo  {journal} {Physical Review Letters}\
  }\textbf {\bibinfo {volume} {105}},\ \bibinfo {pages} {177002} (\bibinfo
  {year} {2010})}\BibitemShut {NoStop}%
\bibitem [{\citenamefont {Lutchyn}\ \emph {et~al.}(2018)\citenamefont
  {Lutchyn}, \citenamefont {Bakkers}, \citenamefont {Kouwenhoven},
  \citenamefont {Krogstrup}, \citenamefont {Marcus},\ and\ \citenamefont
  {Oreg}}]{Lutchyn2018}%
  \BibitemOpen
  \bibfield  {author} {\bibinfo {author} {\bibfnamefont {R.~M.}\ \bibnamefont
  {Lutchyn}}, \bibinfo {author} {\bibfnamefont {E.~P. A.~M.}\ \bibnamefont
  {Bakkers}}, \bibinfo {author} {\bibfnamefont {L.~P.}\ \bibnamefont
  {Kouwenhoven}}, \bibinfo {author} {\bibfnamefont {P.}~\bibnamefont
  {Krogstrup}}, \bibinfo {author} {\bibfnamefont {C.~M.}\ \bibnamefont
  {Marcus}}, \ and\ \bibinfo {author} {\bibfnamefont {Y.}~\bibnamefont
  {Oreg}},\ }\href@noop {} {\bibfield  {journal} {\bibinfo  {journal} {Nature
  Reviews Materials}\ }\textbf {\bibinfo {volume} {3}},\ \bibinfo {pages} {52}
  (\bibinfo {year} {2018})}\BibitemShut {NoStop}%
\bibitem [{\citenamefont {Stanescu}\ and\ \citenamefont
  {Tewari}(2013)}]{Stanescu2013}%
  \BibitemOpen
  \bibfield  {author} {\bibinfo {author} {\bibfnamefont {T.~D.}\ \bibnamefont
  {Stanescu}}\ and\ \bibinfo {author} {\bibfnamefont {S.}~\bibnamefont
  {Tewari}},\ }\href@noop {} {\bibfield  {journal} {\bibinfo  {journal}
  {Journal of Physics: Condensed Matter}\ }\textbf {\bibinfo {volume} {25}},\
  \bibinfo {pages} {29} (\bibinfo {year} {2013})}\BibitemShut {NoStop}%
\bibitem [{\citenamefont {Leijnse}\ and\ \citenamefont
  {Flensberg}(2012)}]{Leijnse2012}%
  \BibitemOpen
  \bibfield  {author} {\bibinfo {author} {\bibfnamefont {M.}~\bibnamefont
  {Leijnse}}\ and\ \bibinfo {author} {\bibfnamefont {K.}~\bibnamefont
  {Flensberg}},\ }\href@noop {} {\bibfield  {journal} {\bibinfo  {journal}
  {Semiconductor Science and Technology}\ }\textbf {\bibinfo {volume} {27}},\
  \bibinfo {pages} {11} (\bibinfo {year} {2012})}\BibitemShut {NoStop}%
\bibitem [{\citenamefont {Nayak}\ \emph {et~al.}(2008)\citenamefont {Nayak},
  \citenamefont {Simon}, \citenamefont {Stern}, \citenamefont {Freedman},\ and\
  \citenamefont {Das~Sarma}}]{Nayak2008}%
  \BibitemOpen
  \bibfield  {author} {\bibinfo {author} {\bibfnamefont {C.}~\bibnamefont
  {Nayak}}, \bibinfo {author} {\bibfnamefont {S.~H.}\ \bibnamefont {Simon}},
  \bibinfo {author} {\bibfnamefont {A.}~\bibnamefont {Stern}}, \bibinfo
  {author} {\bibfnamefont {M.}~\bibnamefont {Freedman}}, \ and\ \bibinfo
  {author} {\bibfnamefont {S.}~\bibnamefont {Das~Sarma}},\ }\href@noop {}
  {\bibfield  {journal} {\bibinfo  {journal} {Reviews of Modern Physics}\
  }\textbf {\bibinfo {volume} {80}},\ \bibinfo {pages} {1083} (\bibinfo {year}
  {2008})}\BibitemShut {NoStop}%
\bibitem [{\citenamefont {Aasen}\ \emph {et~al.}(2016)\citenamefont {Aasen},
  \citenamefont {Hell}, \citenamefont {Mishmash}, \citenamefont {Higginbotham},
  \citenamefont {Danon}, \citenamefont {Leijnse}, \citenamefont {Jespersen},
  \citenamefont {Folk}, \citenamefont {Marcus}, \citenamefont {Flensberg} \emph
  {et~al.}}]{aasen2016milestones}%
  \BibitemOpen
  \bibfield  {author} {\bibinfo {author} {\bibfnamefont {D.}~\bibnamefont
  {Aasen}}, \bibinfo {author} {\bibfnamefont {M.}~\bibnamefont {Hell}},
  \bibinfo {author} {\bibfnamefont {R.~V.}\ \bibnamefont {Mishmash}}, \bibinfo
  {author} {\bibfnamefont {A.}~\bibnamefont {Higginbotham}}, \bibinfo {author}
  {\bibfnamefont {J.}~\bibnamefont {Danon}}, \bibinfo {author} {\bibfnamefont
  {M.}~\bibnamefont {Leijnse}}, \bibinfo {author} {\bibfnamefont {T.~S.}\
  \bibnamefont {Jespersen}}, \bibinfo {author} {\bibfnamefont {J.~A.}\
  \bibnamefont {Folk}}, \bibinfo {author} {\bibfnamefont {C.~M.}\ \bibnamefont
  {Marcus}}, \bibinfo {author} {\bibfnamefont {K.}~\bibnamefont {Flensberg}},
  \emph {et~al.},\ }\href@noop {} {\bibfield  {journal} {\bibinfo  {journal}
  {Physical Review X}\ }\textbf {\bibinfo {volume} {6}},\ \bibinfo {pages}
  {031016} (\bibinfo {year} {2016})}\BibitemShut {NoStop}%
\bibitem [{\citenamefont {Krogstrup}\ \emph {et~al.}(2015)\citenamefont
  {Krogstrup}, \citenamefont {Ziino}, \citenamefont {Chang}, \citenamefont
  {Albrecht}, \citenamefont {Madsen}, \citenamefont {Johnson}, \citenamefont
  {Nyg{\aa}rd}, \citenamefont {Marcus},\ and\ \citenamefont
  {Jespersen}}]{krogstrup2015epitaxy}%
  \BibitemOpen
  \bibfield  {author} {\bibinfo {author} {\bibfnamefont {P.}~\bibnamefont
  {Krogstrup}}, \bibinfo {author} {\bibfnamefont {N.}~\bibnamefont {Ziino}},
  \bibinfo {author} {\bibfnamefont {W.}~\bibnamefont {Chang}}, \bibinfo
  {author} {\bibfnamefont {S.}~\bibnamefont {Albrecht}}, \bibinfo {author}
  {\bibfnamefont {M.}~\bibnamefont {Madsen}}, \bibinfo {author} {\bibfnamefont
  {E.}~\bibnamefont {Johnson}}, \bibinfo {author} {\bibfnamefont
  {J.}~\bibnamefont {Nyg{\aa}rd}}, \bibinfo {author} {\bibfnamefont
  {C.}~\bibnamefont {Marcus}}, \ and\ \bibinfo {author} {\bibfnamefont
  {T.}~\bibnamefont {Jespersen}},\ }\href@noop {} {\bibfield  {journal}
  {\bibinfo  {journal} {Nature Materials}\ }\textbf {\bibinfo {volume} {14}},\
  \bibinfo {pages} {400} (\bibinfo {year} {2015})}\BibitemShut {NoStop}%
\bibitem [{\citenamefont {Sestoft}\ \emph {et~al.}(2018)\citenamefont
  {Sestoft}, \citenamefont {Kanne}, \citenamefont {Gejl}, \citenamefont {von
  Soosten}, \citenamefont {Yodh}, \citenamefont {Sherman}, \citenamefont
  {Tarasinski}, \citenamefont {Wimmer}, \citenamefont {Johnson}, \citenamefont
  {Deng}, \citenamefont {Nyg{\aa}rd}, \citenamefont {Marcus},\ and\
  \citenamefont {Krogstrup}}]{sestoft2017hybrid}%
  \BibitemOpen
  \bibfield  {author} {\bibinfo {author} {\bibfnamefont {J.~E.}\ \bibnamefont
  {Sestoft}}, \bibinfo {author} {\bibfnamefont {T.}~\bibnamefont {Kanne}},
  \bibinfo {author} {\bibfnamefont {A.~N.}\ \bibnamefont {Gejl}}, \bibinfo
  {author} {\bibfnamefont {M.}~\bibnamefont {von Soosten}}, \bibinfo {author}
  {\bibfnamefont {J.~S.}\ \bibnamefont {Yodh}}, \bibinfo {author}
  {\bibfnamefont {D.}~\bibnamefont {Sherman}}, \bibinfo {author} {\bibfnamefont
  {B.}~\bibnamefont {Tarasinski}}, \bibinfo {author} {\bibfnamefont
  {M.}~\bibnamefont {Wimmer}}, \bibinfo {author} {\bibfnamefont
  {E.}~\bibnamefont {Johnson}}, \bibinfo {author} {\bibfnamefont
  {M.}~\bibnamefont {Deng}}, \bibinfo {author} {\bibfnamefont {J.}~\bibnamefont
  {Nyg{\aa}rd}}, \bibinfo {author} {\bibfnamefont {C.}~\bibnamefont {Marcus}},
  \ and\ \bibinfo {author} {\bibfnamefont {P.}~\bibnamefont {Krogstrup}},\
  }\href@noop {} {\bibfield  {journal} {\bibinfo  {journal} {Physical Review
  Materials}\ }\textbf {\bibinfo {volume} {2}},\ \bibinfo {pages} {044202}
  (\bibinfo {year} {2018})}\BibitemShut {NoStop}%
\bibitem [{\citenamefont {Krizek}\ \emph {et~al.}(2018)\citenamefont {Krizek},
  \citenamefont {Sestoft}, \citenamefont {Aseev}, \citenamefont
  {Marti-Sanchez}, \citenamefont {Vaitiek\ifmmode~\dot{e}\else \.{e}\fi{}nas},
  \citenamefont {Casparis}, \citenamefont {Khan}, \citenamefont {Liu},
  \citenamefont {Stankevi\ifmmode~\check{c}\else \v{c}\fi{}}, \citenamefont
  {Whiticar}, \citenamefont {Fursina}, \citenamefont {Boekhout}, \citenamefont
  {Koops}, \citenamefont {Uccelli}, \citenamefont {Kouwenhoven}, \citenamefont
  {Marcus}, \citenamefont {Arbiol},\ and\ \citenamefont
  {Krogstrup}}]{KrizekSAG2018}%
  \BibitemOpen
  \bibfield  {author} {\bibinfo {author} {\bibfnamefont {F.}~\bibnamefont
  {Krizek}}, \bibinfo {author} {\bibfnamefont {J.~E.}\ \bibnamefont {Sestoft}},
  \bibinfo {author} {\bibfnamefont {P.}~\bibnamefont {Aseev}}, \bibinfo
  {author} {\bibfnamefont {S.}~\bibnamefont {Marti-Sanchez}}, \bibinfo {author}
  {\bibfnamefont {S.}~\bibnamefont {Vaitiek\ifmmode~\dot{e}\else
  \.{e}\fi{}nas}}, \bibinfo {author} {\bibfnamefont {L.}~\bibnamefont
  {Casparis}}, \bibinfo {author} {\bibfnamefont {S.~A.}\ \bibnamefont {Khan}},
  \bibinfo {author} {\bibfnamefont {Y.}~\bibnamefont {Liu}}, \bibinfo {author}
  {\bibfnamefont {T.~c.~v.}\ \bibnamefont {Stankevi\ifmmode~\check{c}\else
  \v{c}\fi{}}}, \bibinfo {author} {\bibfnamefont {A.~M.}\ \bibnamefont
  {Whiticar}}, \bibinfo {author} {\bibfnamefont {A.}~\bibnamefont {Fursina}},
  \bibinfo {author} {\bibfnamefont {F.}~\bibnamefont {Boekhout}}, \bibinfo
  {author} {\bibfnamefont {R.}~\bibnamefont {Koops}}, \bibinfo {author}
  {\bibfnamefont {E.}~\bibnamefont {Uccelli}}, \bibinfo {author} {\bibfnamefont
  {L.~P.}\ \bibnamefont {Kouwenhoven}}, \bibinfo {author} {\bibfnamefont
  {C.~M.}\ \bibnamefont {Marcus}}, \bibinfo {author} {\bibfnamefont
  {J.}~\bibnamefont {Arbiol}}, \ and\ \bibinfo {author} {\bibfnamefont
  {P.}~\bibnamefont {Krogstrup}},\ }\href@noop {} {\bibfield  {journal}
  {\bibinfo  {journal} {Physical Review. Materials}\ }\textbf {\bibinfo
  {volume} {2}},\ \bibinfo {pages} {093401} (\bibinfo {year}
  {2018})}\BibitemShut {NoStop}%
\bibitem [{\citenamefont {Liu}\ \emph {et~al.}(2019)\citenamefont {Liu},
  \citenamefont {Vaitiekenas}, \citenamefont {Mart{\'\i}-S{\'a}nchez},
  \citenamefont {Koch}, \citenamefont {Hart}, \citenamefont {Cui},
  \citenamefont {Kanne}, \citenamefont {Khan}, \citenamefont {Tanta},
  \citenamefont {Upadhyay} \emph {et~al.}}]{liu2019semiconductor}%
  \BibitemOpen
  \bibfield  {author} {\bibinfo {author} {\bibfnamefont {Y.}~\bibnamefont
  {Liu}}, \bibinfo {author} {\bibfnamefont {S.}~\bibnamefont {Vaitiekenas}},
  \bibinfo {author} {\bibfnamefont {S.}~\bibnamefont {Mart{\'\i}-S{\'a}nchez}},
  \bibinfo {author} {\bibfnamefont {C.}~\bibnamefont {Koch}}, \bibinfo {author}
  {\bibfnamefont {S.}~\bibnamefont {Hart}}, \bibinfo {author} {\bibfnamefont
  {Z.}~\bibnamefont {Cui}}, \bibinfo {author} {\bibfnamefont {T.}~\bibnamefont
  {Kanne}}, \bibinfo {author} {\bibfnamefont {S.~A.}\ \bibnamefont {Khan}},
  \bibinfo {author} {\bibfnamefont {R.}~\bibnamefont {Tanta}}, \bibinfo
  {author} {\bibfnamefont {S.}~\bibnamefont {Upadhyay}},  \emph {et~al.},\
  }\href@noop {} {\bibfield  {journal} {\bibinfo  {journal} {Nano Letters}\
  }\textbf {\bibinfo {volume} {20}},\ \bibinfo {pages} {456} (\bibinfo {year}
  {2019})}\BibitemShut {NoStop}%
\bibitem [{\citenamefont {Gill}\ \emph {et~al.}(2018)\citenamefont {Gill},
  \citenamefont {Damasco}, \citenamefont {Janicek}, \citenamefont {Durkin},
  \citenamefont {Humbert}, \citenamefont {Gazibegovic}, \citenamefont {Car},
  \citenamefont {Bakkers}, \citenamefont {Huang},\ and\ \citenamefont
  {Mason}}]{gill2018selective}%
  \BibitemOpen
  \bibfield  {author} {\bibinfo {author} {\bibfnamefont {S.~T.}\ \bibnamefont
  {Gill}}, \bibinfo {author} {\bibfnamefont {J.}~\bibnamefont {Damasco}},
  \bibinfo {author} {\bibfnamefont {B.~E.}\ \bibnamefont {Janicek}}, \bibinfo
  {author} {\bibfnamefont {M.~S.}\ \bibnamefont {Durkin}}, \bibinfo {author}
  {\bibfnamefont {V.}~\bibnamefont {Humbert}}, \bibinfo {author} {\bibfnamefont
  {S.}~\bibnamefont {Gazibegovic}}, \bibinfo {author} {\bibfnamefont
  {D.}~\bibnamefont {Car}}, \bibinfo {author} {\bibfnamefont {E.~P.}\
  \bibnamefont {Bakkers}}, \bibinfo {author} {\bibfnamefont {P.~Y.}\
  \bibnamefont {Huang}}, \ and\ \bibinfo {author} {\bibfnamefont
  {N.}~\bibnamefont {Mason}},\ }\href@noop {} {\bibfield  {journal} {\bibinfo
  {journal} {Nano Letters}\ }\textbf {\bibinfo {volume} {18}},\ \bibinfo
  {pages} {6121} (\bibinfo {year} {2018})}\BibitemShut {NoStop}%
\bibitem [{\citenamefont {Mourik}\ \emph {et~al.}(2012)\citenamefont {Mourik},
  \citenamefont {Zuo}, \citenamefont {Frolov}, \citenamefont {Plissard},
  \citenamefont {Bakkers},\ and\ \citenamefont {Kouwenhoven}}]{Mourik2012}%
  \BibitemOpen
  \bibfield  {author} {\bibinfo {author} {\bibfnamefont {V.}~\bibnamefont
  {Mourik}}, \bibinfo {author} {\bibfnamefont {K.}~\bibnamefont {Zuo}},
  \bibinfo {author} {\bibfnamefont {S.~M.}\ \bibnamefont {Frolov}}, \bibinfo
  {author} {\bibfnamefont {S.~R.}\ \bibnamefont {Plissard}}, \bibinfo {author}
  {\bibfnamefont {E.~P. A.~M.}\ \bibnamefont {Bakkers}}, \ and\ \bibinfo
  {author} {\bibfnamefont {L.~P.}\ \bibnamefont {Kouwenhoven}},\ }\href@noop {}
  {\bibfield  {journal} {\bibinfo  {journal} {Science}\ }\textbf {\bibinfo
  {volume} {336}},\ \bibinfo {pages} {1003} (\bibinfo {year}
  {2012})}\BibitemShut {NoStop}%
\bibitem [{\citenamefont {Das}\ \emph {et~al.}(2012)\citenamefont {Das},
  \citenamefont {Ronen}, \citenamefont {Most}, \citenamefont {Oreg},
  \citenamefont {Heiblum},\ and\ \citenamefont {Shtrikman}}]{Das2012}%
  \BibitemOpen
  \bibfield  {author} {\bibinfo {author} {\bibfnamefont {A.}~\bibnamefont
  {Das}}, \bibinfo {author} {\bibfnamefont {Y.}~\bibnamefont {Ronen}}, \bibinfo
  {author} {\bibfnamefont {Y.}~\bibnamefont {Most}}, \bibinfo {author}
  {\bibfnamefont {Y.}~\bibnamefont {Oreg}}, \bibinfo {author} {\bibfnamefont
  {M.}~\bibnamefont {Heiblum}}, \ and\ \bibinfo {author} {\bibfnamefont
  {H.}~\bibnamefont {Shtrikman}},\ }\href@noop {} {\bibfield  {journal}
  {\bibinfo  {journal} {Nature Physics}\ }\textbf {\bibinfo {volume} {8}},\
  \bibinfo {pages} {887} (\bibinfo {year} {2012})}\BibitemShut {NoStop}%
\bibitem [{\citenamefont {Rokhinson}\ \emph {et~al.}(2012)\citenamefont
  {Rokhinson}, \citenamefont {Liu},\ and\ \citenamefont
  {Furdyna}}]{Rokhinson2012}%
  \BibitemOpen
  \bibfield  {author} {\bibinfo {author} {\bibfnamefont {L.~P.}\ \bibnamefont
  {Rokhinson}}, \bibinfo {author} {\bibfnamefont {X.}~\bibnamefont {Liu}}, \
  and\ \bibinfo {author} {\bibfnamefont {J.~K.}\ \bibnamefont {Furdyna}},\
  }\href@noop {} {\bibfield  {journal} {\bibinfo  {journal} {Nature Physics}\
  }\textbf {\bibinfo {volume} {8}},\ \bibinfo {pages} {795} (\bibinfo {year}
  {2012})}\BibitemShut {NoStop}%
\bibitem [{\citenamefont {Deng}\ \emph {et~al.}(2012)\citenamefont {Deng},
  \citenamefont {Yu}, \citenamefont {Huang}, \citenamefont {Larsson},
  \citenamefont {Caroff},\ and\ \citenamefont {Xu}}]{deng2012anomalous}%
  \BibitemOpen
  \bibfield  {author} {\bibinfo {author} {\bibfnamefont {M.}~\bibnamefont
  {Deng}}, \bibinfo {author} {\bibfnamefont {C.}~\bibnamefont {Yu}}, \bibinfo
  {author} {\bibfnamefont {G.}~\bibnamefont {Huang}}, \bibinfo {author}
  {\bibfnamefont {M.}~\bibnamefont {Larsson}}, \bibinfo {author} {\bibfnamefont
  {P.}~\bibnamefont {Caroff}}, \ and\ \bibinfo {author} {\bibfnamefont
  {H.}~\bibnamefont {Xu}},\ }\href@noop {} {\bibfield  {journal} {\bibinfo
  {journal} {Nano Letters}\ }\textbf {\bibinfo {volume} {12}},\ \bibinfo
  {pages} {6414} (\bibinfo {year} {2012})}\BibitemShut {NoStop}%
\bibitem [{\citenamefont {Finck}\ \emph {et~al.}(2013)\citenamefont {Finck},
  \citenamefont {Van~Harlingen}, \citenamefont {Mohseni}, \citenamefont
  {Jung},\ and\ \citenamefont {Li}}]{Finck2013}%
  \BibitemOpen
  \bibfield  {author} {\bibinfo {author} {\bibfnamefont {A.~D.~K.}\
  \bibnamefont {Finck}}, \bibinfo {author} {\bibfnamefont {D.~J.}\ \bibnamefont
  {Van~Harlingen}}, \bibinfo {author} {\bibfnamefont {P.~K.}\ \bibnamefont
  {Mohseni}}, \bibinfo {author} {\bibfnamefont {K.}~\bibnamefont {Jung}}, \
  and\ \bibinfo {author} {\bibfnamefont {X.}~\bibnamefont {Li}},\ }\href@noop
  {} {\bibfield  {journal} {\bibinfo  {journal} {Physical Review Letters}\
  }\textbf {\bibinfo {volume} {110}},\ \bibinfo {pages} {126406} (\bibinfo
  {year} {2013})}\BibitemShut {NoStop}%
\bibitem [{\citenamefont {Deng}\ \emph {et~al.}(2016)\citenamefont {Deng},
  \citenamefont {Vaitiekenas}, \citenamefont {Hansen}, \citenamefont {Danon},
  \citenamefont {Leijnse}, \citenamefont {Flensberg}, \citenamefont {Nygard},
  \citenamefont {Krogstrup},\ and\ \citenamefont {Marcus}}]{Deng2016}%
  \BibitemOpen
  \bibfield  {author} {\bibinfo {author} {\bibfnamefont {M.~T.}\ \bibnamefont
  {Deng}}, \bibinfo {author} {\bibfnamefont {S.}~\bibnamefont {Vaitiekenas}},
  \bibinfo {author} {\bibfnamefont {E.~B.}\ \bibnamefont {Hansen}}, \bibinfo
  {author} {\bibfnamefont {J.}~\bibnamefont {Danon}}, \bibinfo {author}
  {\bibfnamefont {M.}~\bibnamefont {Leijnse}}, \bibinfo {author} {\bibfnamefont
  {K.}~\bibnamefont {Flensberg}}, \bibinfo {author} {\bibfnamefont
  {J.}~\bibnamefont {Nygard}}, \bibinfo {author} {\bibfnamefont
  {P.}~\bibnamefont {Krogstrup}}, \ and\ \bibinfo {author} {\bibfnamefont
  {C.~M.}\ \bibnamefont {Marcus}},\ }\href@noop {} {\bibfield  {journal}
  {\bibinfo  {journal} {Science}\ }\textbf {\bibinfo {volume} {354}},\ \bibinfo
  {pages} {1557} (\bibinfo {year} {2016})}\BibitemShut {NoStop}%
\bibitem [{\citenamefont {Zhang}\ \emph {et~al.}(2018)\citenamefont {Zhang},
  \citenamefont {Liu}, \citenamefont {Gazibegovic}, \citenamefont {Xu},
  \citenamefont {Logan}, \citenamefont {Wang}, \citenamefont {Loo},
  \citenamefont {Bommer}, \citenamefont {Moor}, \citenamefont {Car},
  \citenamefont {Veld}, \citenamefont {Veldhoven}, \citenamefont {Koelling},
  \citenamefont {Verheijen}, \citenamefont {Pendharkar}, \citenamefont
  {Pennachio}, \citenamefont {Shojaei}, \citenamefont {Lee}, \citenamefont
  {Palmstrom}, \citenamefont {Bakkers}, \citenamefont {Sarma},\ and\
  \citenamefont {Kouwenhoven}}]{Zhang2018}%
  \BibitemOpen
  \bibfield  {author} {\bibinfo {author} {\bibfnamefont {H.}~\bibnamefont
  {Zhang}}, \bibinfo {author} {\bibfnamefont {C.-X.}\ \bibnamefont {Liu}},
  \bibinfo {author} {\bibfnamefont {S.}~\bibnamefont {Gazibegovic}}, \bibinfo
  {author} {\bibfnamefont {D.}~\bibnamefont {Xu}}, \bibinfo {author}
  {\bibfnamefont {J.~A.}\ \bibnamefont {Logan}}, \bibinfo {author}
  {\bibfnamefont {G.}~\bibnamefont {Wang}}, \bibinfo {author} {\bibfnamefont
  {N.~V.}\ \bibnamefont {Loo}}, \bibinfo {author} {\bibfnamefont {J.~D.~S.}\
  \bibnamefont {Bommer}}, \bibinfo {author} {\bibfnamefont {M.~W. A.~d.}\
  \bibnamefont {Moor}}, \bibinfo {author} {\bibfnamefont {D.}~\bibnamefont
  {Car}}, \bibinfo {author} {\bibfnamefont {R.~L. M. O.~h.}\ \bibnamefont
  {Veld}}, \bibinfo {author} {\bibfnamefont {P.~J.~V.}\ \bibnamefont
  {Veldhoven}}, \bibinfo {author} {\bibfnamefont {S.}~\bibnamefont {Koelling}},
  \bibinfo {author} {\bibfnamefont {M.~A.}\ \bibnamefont {Verheijen}}, \bibinfo
  {author} {\bibfnamefont {M.}~\bibnamefont {Pendharkar}}, \bibinfo {author}
  {\bibfnamefont {D.~J.}\ \bibnamefont {Pennachio}}, \bibinfo {author}
  {\bibfnamefont {B.}~\bibnamefont {Shojaei}}, \bibinfo {author} {\bibfnamefont
  {J.~S.}\ \bibnamefont {Lee}}, \bibinfo {author} {\bibfnamefont {C.~J.}\
  \bibnamefont {Palmstrom}}, \bibinfo {author} {\bibfnamefont {E.~P. A.~M.}\
  \bibnamefont {Bakkers}}, \bibinfo {author} {\bibfnamefont {S.~D.}\
  \bibnamefont {Sarma}}, \ and\ \bibinfo {author} {\bibfnamefont {L.~P.}\
  \bibnamefont {Kouwenhoven}},\ }\href@noop {} {\bibfield  {journal} {\bibinfo
  {journal} {Nature}\ }\textbf {\bibinfo {volume} {556}},\ \bibinfo {pages}
  {74} (\bibinfo {year} {2018})}\BibitemShut {NoStop}%
\bibitem [{\citenamefont {Chen}\ \emph {et~al.}(2017)\citenamefont {Chen},
  \citenamefont {Yu}, \citenamefont {Stenger}, \citenamefont {Hocevar},
  \citenamefont {Car}, \citenamefont {Plissard}, \citenamefont {Bakkers},
  \citenamefont {Stanescu},\ and\ \citenamefont {Frolov}}]{Chen2017}%
  \BibitemOpen
  \bibfield  {author} {\bibinfo {author} {\bibfnamefont {J.}~\bibnamefont
  {Chen}}, \bibinfo {author} {\bibfnamefont {P.}~\bibnamefont {Yu}}, \bibinfo
  {author} {\bibfnamefont {J.}~\bibnamefont {Stenger}}, \bibinfo {author}
  {\bibfnamefont {M.}~\bibnamefont {Hocevar}}, \bibinfo {author} {\bibfnamefont
  {D.}~\bibnamefont {Car}}, \bibinfo {author} {\bibfnamefont {S.~R.}\
  \bibnamefont {Plissard}}, \bibinfo {author} {\bibfnamefont {E.~P. A.~M.}\
  \bibnamefont {Bakkers}}, \bibinfo {author} {\bibfnamefont {T.~D.}\
  \bibnamefont {Stanescu}}, \ and\ \bibinfo {author} {\bibfnamefont {S.~M.}\
  \bibnamefont {Frolov}},\ }\href@noop {} {\bibfield  {journal} {\bibinfo
  {journal} {Science Advances}\ }\textbf {\bibinfo {volume} {3}},\ \bibinfo
  {pages} {e1701476} (\bibinfo {year} {2017})}\BibitemShut {NoStop}%
\bibitem [{\citenamefont {Gul}\ \emph {et~al.}(2018)\citenamefont {Gul},
  \citenamefont {Zhang}, \citenamefont {Bommer}, \citenamefont {Moor},
  \citenamefont {Car}, \citenamefont {Plissard}, \citenamefont {Bakkers},
  \citenamefont {Geresdi}, \citenamefont {Watanabe}, \citenamefont
  {Taniguchi},\ and\ \citenamefont {Kouwenhoven}}]{Gul2018}%
  \BibitemOpen
  \bibfield  {author} {\bibinfo {author} {\bibfnamefont {O.}~\bibnamefont
  {Gul}}, \bibinfo {author} {\bibfnamefont {H.}~\bibnamefont {Zhang}}, \bibinfo
  {author} {\bibfnamefont {J.~D.~S.}\ \bibnamefont {Bommer}}, \bibinfo {author}
  {\bibfnamefont {M.~W. A.~d.}\ \bibnamefont {Moor}}, \bibinfo {author}
  {\bibfnamefont {D.}~\bibnamefont {Car}}, \bibinfo {author} {\bibfnamefont
  {S.~R.}\ \bibnamefont {Plissard}}, \bibinfo {author} {\bibfnamefont {E.~P.
  A.~M.}\ \bibnamefont {Bakkers}}, \bibinfo {author} {\bibfnamefont
  {A.}~\bibnamefont {Geresdi}}, \bibinfo {author} {\bibfnamefont
  {K.}~\bibnamefont {Watanabe}}, \bibinfo {author} {\bibfnamefont
  {T.}~\bibnamefont {Taniguchi}}, \ and\ \bibinfo {author} {\bibfnamefont
  {L.~P.}\ \bibnamefont {Kouwenhoven}},\ }\href@noop {} {\bibfield  {journal}
  {\bibinfo  {journal} {Nature Nanotechnology}\ }\textbf {\bibinfo {volume}
  {13}},\ \bibinfo {pages} {192} (\bibinfo {year} {2018})}\BibitemShut
  {NoStop}%
\bibitem [{\citenamefont {Liu}\ \emph {et~al.}(2018)\citenamefont {Liu},
  \citenamefont {Sau},\ and\ \citenamefont {Das~Sarma}}]{PhysRevB.97.214502}%
  \BibitemOpen
  \bibfield  {author} {\bibinfo {author} {\bibfnamefont {C.-X.}\ \bibnamefont
  {Liu}}, \bibinfo {author} {\bibfnamefont {J.~D.}\ \bibnamefont {Sau}}, \ and\
  \bibinfo {author} {\bibfnamefont {S.}~\bibnamefont {Das~Sarma}},\ }\href@noop
  {} {\bibfield  {journal} {\bibinfo  {journal} {Physical Review B}\ }\textbf
  {\bibinfo {volume} {97}},\ \bibinfo {pages} {214502} (\bibinfo {year}
  {2018})}\BibitemShut {NoStop}%
\bibitem [{\citenamefont {Lee}\ \emph {et~al.}(2014)\citenamefont {Lee},
  \citenamefont {Jiang}, \citenamefont {Houzet}, \citenamefont {Aguado},
  \citenamefont {Lieber},\ and\ \citenamefont {De~Franceschi}}]{lee2014spin}%
  \BibitemOpen
  \bibfield  {author} {\bibinfo {author} {\bibfnamefont {E.~J.}\ \bibnamefont
  {Lee}}, \bibinfo {author} {\bibfnamefont {X.}~\bibnamefont {Jiang}}, \bibinfo
  {author} {\bibfnamefont {M.}~\bibnamefont {Houzet}}, \bibinfo {author}
  {\bibfnamefont {R.}~\bibnamefont {Aguado}}, \bibinfo {author} {\bibfnamefont
  {C.~M.}\ \bibnamefont {Lieber}}, \ and\ \bibinfo {author} {\bibfnamefont
  {S.}~\bibnamefont {De~Franceschi}},\ }\href@noop {} {\bibfield  {journal}
  {\bibinfo  {journal} {Nature Nanotechnology}\ }\textbf {\bibinfo {volume}
  {9}},\ \bibinfo {pages} {79} (\bibinfo {year} {2014})}\BibitemShut {NoStop}%
\bibitem [{\citenamefont {Pan}\ and\ \citenamefont
  {Sarma}(2019)}]{pan2019zerobias}%
  \BibitemOpen
  \bibfield  {author} {\bibinfo {author} {\bibfnamefont {H.}~\bibnamefont
  {Pan}}\ and\ \bibinfo {author} {\bibfnamefont {S.~D.}\ \bibnamefont
  {Sarma}},\ }\href@noop {} {\bibfield  {journal} {\bibinfo  {journal}
  {arXiv:1910.11413}\ } (\bibinfo {year} {2019})}\BibitemShut {NoStop}%
\bibitem [{\citenamefont {Albrecht}\ \emph {et~al.}(2016)\citenamefont
  {Albrecht}, \citenamefont {Higginbotham}, \citenamefont {Madsen},
  \citenamefont {Kuemmeth}, \citenamefont {Jespersen}, \citenamefont {Nygard},
  \citenamefont {Krogstrup},\ and\ \citenamefont {Marcus}}]{Albrecht2016}%
  \BibitemOpen
  \bibfield  {author} {\bibinfo {author} {\bibfnamefont {S.~M.}\ \bibnamefont
  {Albrecht}}, \bibinfo {author} {\bibfnamefont {A.~P.}\ \bibnamefont
  {Higginbotham}}, \bibinfo {author} {\bibfnamefont {M.}~\bibnamefont
  {Madsen}}, \bibinfo {author} {\bibfnamefont {F.}~\bibnamefont {Kuemmeth}},
  \bibinfo {author} {\bibfnamefont {T.~S.}\ \bibnamefont {Jespersen}}, \bibinfo
  {author} {\bibfnamefont {J.}~\bibnamefont {Nygard}}, \bibinfo {author}
  {\bibfnamefont {P.}~\bibnamefont {Krogstrup}}, \ and\ \bibinfo {author}
  {\bibfnamefont {C.~M.}\ \bibnamefont {Marcus}},\ }\href@noop {} {\bibfield
  {journal} {\bibinfo  {journal} {Nature}\ }\textbf {\bibinfo {volume} {531}},\
  \bibinfo {pages} {206} (\bibinfo {year} {2016})}\BibitemShut {NoStop}%
\bibitem [{\citenamefont {Nichele}\ \emph {et~al.}(2017)\citenamefont
  {Nichele}, \citenamefont {Drachmann}, \citenamefont {Whiticar}, \citenamefont
  {O'Farrell}, \citenamefont {Suominen}, \citenamefont {Fornieri},
  \citenamefont {Wang}, \citenamefont {Gardner}, \citenamefont {Thomas},
  \citenamefont {Hatke}, \citenamefont {Krogstrup}, \citenamefont {Manfra},
  \citenamefont {Flensberg},\ and\ \citenamefont {Marcus}}]{Nichele2017}%
  \BibitemOpen
  \bibfield  {author} {\bibinfo {author} {\bibfnamefont {F.}~\bibnamefont
  {Nichele}}, \bibinfo {author} {\bibfnamefont {A.~C.~C.}\ \bibnamefont
  {Drachmann}}, \bibinfo {author} {\bibfnamefont {A.~M.}\ \bibnamefont
  {Whiticar}}, \bibinfo {author} {\bibfnamefont {E.~C.~T.}\ \bibnamefont
  {O'Farrell}}, \bibinfo {author} {\bibfnamefont {H.~J.}\ \bibnamefont
  {Suominen}}, \bibinfo {author} {\bibfnamefont {A.}~\bibnamefont {Fornieri}},
  \bibinfo {author} {\bibfnamefont {T.}~\bibnamefont {Wang}}, \bibinfo {author}
  {\bibfnamefont {G.~C.}\ \bibnamefont {Gardner}}, \bibinfo {author}
  {\bibfnamefont {C.}~\bibnamefont {Thomas}}, \bibinfo {author} {\bibfnamefont
  {A.~T.}\ \bibnamefont {Hatke}}, \bibinfo {author} {\bibfnamefont
  {P.}~\bibnamefont {Krogstrup}}, \bibinfo {author} {\bibfnamefont {M.~J.}\
  \bibnamefont {Manfra}}, \bibinfo {author} {\bibfnamefont {K.}~\bibnamefont
  {Flensberg}}, \ and\ \bibinfo {author} {\bibfnamefont {C.~M.}\ \bibnamefont
  {Marcus}},\ }\href@noop {} {\bibfield  {journal} {\bibinfo  {journal}
  {Physical Review Letters}\ }\textbf {\bibinfo {volume} {119}},\ \bibinfo
  {pages} {136803} (\bibinfo {year} {2017})}\BibitemShut {NoStop}%
\bibitem [{\citenamefont {Krizek}\ \emph {et~al.}(2017)\citenamefont {Krizek},
  \citenamefont {Kanne}, \citenamefont {Razmadze}, \citenamefont {Johnson},
  \citenamefont {Nyg{\aa}rd}, \citenamefont {Marcus},\ and\ \citenamefont
  {Krogstrup}}]{krizek2017growth}%
  \BibitemOpen
  \bibfield  {author} {\bibinfo {author} {\bibfnamefont {F.}~\bibnamefont
  {Krizek}}, \bibinfo {author} {\bibfnamefont {T.}~\bibnamefont {Kanne}},
  \bibinfo {author} {\bibfnamefont {D.}~\bibnamefont {Razmadze}}, \bibinfo
  {author} {\bibfnamefont {E.}~\bibnamefont {Johnson}}, \bibinfo {author}
  {\bibfnamefont {J.}~\bibnamefont {Nyg{\aa}rd}}, \bibinfo {author}
  {\bibfnamefont {C.~M.}\ \bibnamefont {Marcus}}, \ and\ \bibinfo {author}
  {\bibfnamefont {P.}~\bibnamefont {Krogstrup}},\ }\href@noop {} {\bibfield
  {journal} {\bibinfo  {journal} {Nano Letters}\ }\textbf {\bibinfo {volume}
  {17}},\ \bibinfo {pages} {6090} (\bibinfo {year} {2017})}\BibitemShut
  {NoStop}%
\bibitem [{\citenamefont {Gazibegovic}\ \emph {et~al.}(2017)\citenamefont
  {Gazibegovic}, \citenamefont {Car}, \citenamefont {Zhang}, \citenamefont
  {Balk}, \citenamefont {Logan}, \citenamefont {de~Moor}, \citenamefont
  {Cassidy}, \citenamefont {Schmits}, \citenamefont {Xu}, \citenamefont {Wang}
  \emph {et~al.}}]{gazibegovic2017epitaxy}%
  \BibitemOpen
  \bibfield  {author} {\bibinfo {author} {\bibfnamefont {S.}~\bibnamefont
  {Gazibegovic}}, \bibinfo {author} {\bibfnamefont {D.}~\bibnamefont {Car}},
  \bibinfo {author} {\bibfnamefont {H.}~\bibnamefont {Zhang}}, \bibinfo
  {author} {\bibfnamefont {S.~C.}\ \bibnamefont {Balk}}, \bibinfo {author}
  {\bibfnamefont {J.~A.}\ \bibnamefont {Logan}}, \bibinfo {author}
  {\bibfnamefont {M.~W.}\ \bibnamefont {de~Moor}}, \bibinfo {author}
  {\bibfnamefont {M.~C.}\ \bibnamefont {Cassidy}}, \bibinfo {author}
  {\bibfnamefont {R.}~\bibnamefont {Schmits}}, \bibinfo {author} {\bibfnamefont
  {D.}~\bibnamefont {Xu}}, \bibinfo {author} {\bibfnamefont {G.}~\bibnamefont
  {Wang}},  \emph {et~al.},\ }\href@noop {} {\bibfield  {journal} {\bibinfo
  {journal} {Nature}\ }\textbf {\bibinfo {volume} {548}},\ \bibinfo {pages}
  {434} (\bibinfo {year} {2017})}\BibitemShut {NoStop}%
\bibitem [{\citenamefont {Carrad}\ \emph {et~al.}(2019)\citenamefont {Carrad},
  \citenamefont {Bjergfelt}, \citenamefont {Kanne}, \citenamefont {Aagesen},
  \citenamefont {Krizek}, \citenamefont {Fiordaliso}, \citenamefont {Johnson},
  \citenamefont {Nyg{\aa}rd},\ and\ \citenamefont
  {Jespersen}}]{carrad2019shadow}%
  \BibitemOpen
  \bibfield  {author} {\bibinfo {author} {\bibfnamefont {D.~J.}\ \bibnamefont
  {Carrad}}, \bibinfo {author} {\bibfnamefont {M.}~\bibnamefont {Bjergfelt}},
  \bibinfo {author} {\bibfnamefont {T.}~\bibnamefont {Kanne}}, \bibinfo
  {author} {\bibfnamefont {M.}~\bibnamefont {Aagesen}}, \bibinfo {author}
  {\bibfnamefont {F.}~\bibnamefont {Krizek}}, \bibinfo {author} {\bibfnamefont
  {E.~M.}\ \bibnamefont {Fiordaliso}}, \bibinfo {author} {\bibfnamefont
  {E.}~\bibnamefont {Johnson}}, \bibinfo {author} {\bibfnamefont
  {J.}~\bibnamefont {Nyg{\aa}rd}}, \ and\ \bibinfo {author} {\bibfnamefont
  {T.~S.}\ \bibnamefont {Jespersen}},\ }\href@noop {} {\bibfield  {journal}
  {\bibinfo  {journal} {arXiv preprint arXiv:1911.00460}\ } (\bibinfo {year}
  {2019})}\BibitemShut {NoStop}%
\bibitem [{\citenamefont {Dalacu}\ \emph {et~al.}(2013)\citenamefont {Dalacu},
  \citenamefont {Kam}, \citenamefont {Austing},\ and\ \citenamefont
  {Poole}}]{Dalacu2013}%
  \BibitemOpen
  \bibfield  {author} {\bibinfo {author} {\bibfnamefont {D.}~\bibnamefont
  {Dalacu}}, \bibinfo {author} {\bibfnamefont {A.}~\bibnamefont {Kam}},
  \bibinfo {author} {\bibfnamefont {D.~G.}\ \bibnamefont {Austing}}, \ and\
  \bibinfo {author} {\bibfnamefont {P.~J.}\ \bibnamefont {Poole}},\ }\href@noop
  {} {\bibfield  {journal} {\bibinfo  {journal} {Nano Letters}\ }\textbf
  {\bibinfo {volume} {13}},\ \bibinfo {pages} {2676} (\bibinfo {year}
  {2013})}\BibitemShut {NoStop}%
\bibitem [{\citenamefont {Caroff}\ \emph {et~al.}(2008)\citenamefont {Caroff},
  \citenamefont {Wagner}, \citenamefont {Dick}, \citenamefont {Nilsson},
  \citenamefont {Jeppsson}, \citenamefont {Knut}, \citenamefont {Samuelson},
  \citenamefont {Wallenberg},\ and\ \citenamefont {Wernersson}}]{Caroff2008}%
  \BibitemOpen
  \bibfield  {author} {\bibinfo {author} {\bibfnamefont {P.}~\bibnamefont
  {Caroff}}, \bibinfo {author} {\bibfnamefont {J.~B.}\ \bibnamefont {Wagner}},
  \bibinfo {author} {\bibfnamefont {K.~A.}\ \bibnamefont {Dick}}, \bibinfo
  {author} {\bibfnamefont {H.~A.}\ \bibnamefont {Nilsson}}, \bibinfo {author}
  {\bibfnamefont {M.}~\bibnamefont {Jeppsson}}, \bibinfo {author}
  {\bibfnamefont {D.}~\bibnamefont {Knut}}, \bibinfo {author} {\bibfnamefont
  {L.}~\bibnamefont {Samuelson}}, \bibinfo {author} {\bibfnamefont {L.~R.}\
  \bibnamefont {Wallenberg}}, \ and\ \bibinfo {author} {\bibfnamefont {L.-E.}\
  \bibnamefont {Wernersson}},\ }\href@noop {} {\bibfield  {journal} {\bibinfo
  {journal} {Small}\ }\textbf {\bibinfo {volume} {4}},\ \bibinfo {pages} {878}
  (\bibinfo {year} {2008})}\BibitemShut {NoStop}%
\bibitem [{\citenamefont {Pendharkar}\ \emph {et~al.}(2019)\citenamefont
  {Pendharkar}, \citenamefont {Zhang}, \citenamefont {Wu}, \citenamefont
  {Zarassi}, \citenamefont {Zhang}, \citenamefont {Dempsey}, \citenamefont
  {Lee}, \citenamefont {Harrington}, \citenamefont {Badawy}, \citenamefont
  {Gazibegovic} \emph {et~al.}}]{pendharkar2019parity}%
  \BibitemOpen
  \bibfield  {author} {\bibinfo {author} {\bibfnamefont {M.}~\bibnamefont
  {Pendharkar}}, \bibinfo {author} {\bibfnamefont {B.}~\bibnamefont {Zhang}},
  \bibinfo {author} {\bibfnamefont {H.}~\bibnamefont {Wu}}, \bibinfo {author}
  {\bibfnamefont {A.}~\bibnamefont {Zarassi}}, \bibinfo {author} {\bibfnamefont
  {P.}~\bibnamefont {Zhang}}, \bibinfo {author} {\bibfnamefont
  {C.}~\bibnamefont {Dempsey}}, \bibinfo {author} {\bibfnamefont
  {J.}~\bibnamefont {Lee}}, \bibinfo {author} {\bibfnamefont {S.}~\bibnamefont
  {Harrington}}, \bibinfo {author} {\bibfnamefont {G.}~\bibnamefont {Badawy}},
  \bibinfo {author} {\bibfnamefont {S.}~\bibnamefont {Gazibegovic}},  \emph
  {et~al.},\ }\href@noop {} {\bibfield  {journal} {\bibinfo  {journal} {arXiv
  preprint arXiv:1912.06071}\ } (\bibinfo {year} {2019})}\BibitemShut {NoStop}%
\bibitem [{\citenamefont {Winkler}\ \emph {et~al.}(2016)\citenamefont
  {Winkler}, \citenamefont {Wu}, \citenamefont {Troyer}, \citenamefont
  {Krogstrup},\ and\ \citenamefont {Soluyanov}}]{winkler2016}%
  \BibitemOpen
  \bibfield  {author} {\bibinfo {author} {\bibfnamefont {G.~W.}\ \bibnamefont
  {Winkler}}, \bibinfo {author} {\bibfnamefont {Q.}~\bibnamefont {Wu}},
  \bibinfo {author} {\bibfnamefont {M.}~\bibnamefont {Troyer}}, \bibinfo
  {author} {\bibfnamefont {P.}~\bibnamefont {Krogstrup}}, \ and\ \bibinfo
  {author} {\bibfnamefont {A.~A.}\ \bibnamefont {Soluyanov}},\ }\href@noop {}
  {\bibfield  {journal} {\bibinfo  {journal} {Physical Review Letters}\
  }\textbf {\bibinfo {volume} {117}},\ \bibinfo {pages} {076403} (\bibinfo
  {year} {2016})}\BibitemShut {NoStop}%
\bibitem [{\citenamefont {Potts}\ \emph {et~al.}(2016)\citenamefont {Potts},
  \citenamefont {Friedl}, \citenamefont {Amaduzzi}, \citenamefont {Tang},
  \citenamefont {T\"ut\"unc\"uoglu}, \citenamefont {Matteini}, \citenamefont
  {Alarcon~Llad{\'o}}, \citenamefont {McIntyre},\ and\ \citenamefont
  {Fontcuberta~i Morral}}]{potts2016twinning}%
  \BibitemOpen
  \bibfield  {author} {\bibinfo {author} {\bibfnamefont {H.}~\bibnamefont
  {Potts}}, \bibinfo {author} {\bibfnamefont {M.}~\bibnamefont {Friedl}},
  \bibinfo {author} {\bibfnamefont {F.}~\bibnamefont {Amaduzzi}}, \bibinfo
  {author} {\bibfnamefont {K.}~\bibnamefont {Tang}}, \bibinfo {author}
  {\bibfnamefont {G.}~\bibnamefont {T\"ut\"unc\"uoglu}}, \bibinfo {author}
  {\bibfnamefont {F.}~\bibnamefont {Matteini}}, \bibinfo {author}
  {\bibfnamefont {E.}~\bibnamefont {Alarcon~Llad{\'o}}}, \bibinfo {author}
  {\bibfnamefont {P.~C.}\ \bibnamefont {McIntyre}}, \ and\ \bibinfo {author}
  {\bibfnamefont {A.}~\bibnamefont {Fontcuberta~i Morral}},\ }\href@noop {}
  {\bibfield  {journal} {\bibinfo  {journal} {Nano Letters}\ }\textbf {\bibinfo
  {volume} {16}},\ \bibinfo {pages} {637} (\bibinfo {year} {2016})}\BibitemShut
  {NoStop}%
\bibitem [{\citenamefont {Vegard}(1921)}]{vegard1921konstitution}%
  \BibitemOpen
  \bibfield  {author} {\bibinfo {author} {\bibfnamefont {L.}~\bibnamefont
  {Vegard}},\ }\href@noop {} {\bibfield  {journal} {\bibinfo  {journal}
  {Zeitschrift f{\"u}r Physik}\ }\textbf {\bibinfo {volume} {5}},\ \bibinfo
  {pages} {17} (\bibinfo {year} {1921})}\BibitemShut {NoStop}%
\bibitem [{\citenamefont {Chang}\ \emph {et~al.}(2015)\citenamefont {Chang},
  \citenamefont {Albrecht}, \citenamefont {Jespersen}, \citenamefont
  {Kuemmeth}, \citenamefont {Krogstrup}, \citenamefont {Nyg{\aa}rd},\ and\
  \citenamefont {Marcus}}]{chang2015hard}%
  \BibitemOpen
  \bibfield  {author} {\bibinfo {author} {\bibfnamefont {W.}~\bibnamefont
  {Chang}}, \bibinfo {author} {\bibfnamefont {S.}~\bibnamefont {Albrecht}},
  \bibinfo {author} {\bibfnamefont {T.}~\bibnamefont {Jespersen}}, \bibinfo
  {author} {\bibfnamefont {F.}~\bibnamefont {Kuemmeth}}, \bibinfo {author}
  {\bibfnamefont {P.}~\bibnamefont {Krogstrup}}, \bibinfo {author}
  {\bibfnamefont {J.}~\bibnamefont {Nyg{\aa}rd}}, \ and\ \bibinfo {author}
  {\bibfnamefont {C.}~\bibnamefont {Marcus}},\ }\href@noop {} {\bibfield
  {journal} {\bibinfo  {journal} {Nature Nanotechnology}\ }\textbf {\bibinfo
  {volume} {10}},\ \bibinfo {pages} {232} (\bibinfo {year} {2015})}\BibitemShut
  {NoStop}%
\bibitem [{\citenamefont {Krogstrup}\ \emph {et~al.}(2013)\citenamefont
  {Krogstrup}, \citenamefont {J{\o}rgensen}, \citenamefont {Johnson},
  \citenamefont {Madsen}, \citenamefont {S{\o}rensen}, \citenamefont
  {i~Morral}, \citenamefont {Aagesen}, \citenamefont {Nyg{\aa}rd},\ and\
  \citenamefont {Glas}}]{krogstrup2013advances}%
  \BibitemOpen
  \bibfield  {author} {\bibinfo {author} {\bibfnamefont {P.}~\bibnamefont
  {Krogstrup}}, \bibinfo {author} {\bibfnamefont {H.~I.}\ \bibnamefont
  {J{\o}rgensen}}, \bibinfo {author} {\bibfnamefont {E.}~\bibnamefont
  {Johnson}}, \bibinfo {author} {\bibfnamefont {M.~H.}\ \bibnamefont {Madsen}},
  \bibinfo {author} {\bibfnamefont {C.~B.}\ \bibnamefont {S{\o}rensen}},
  \bibinfo {author} {\bibfnamefont {A.~F.}\ \bibnamefont {i~Morral}}, \bibinfo
  {author} {\bibfnamefont {M.}~\bibnamefont {Aagesen}}, \bibinfo {author}
  {\bibfnamefont {J.}~\bibnamefont {Nyg{\aa}rd}}, \ and\ \bibinfo {author}
  {\bibfnamefont {F.}~\bibnamefont {Glas}},\ }\href@noop {} {\bibfield
  {journal} {\bibinfo  {journal} {Journal of Physics D: Applied Physics}\
  }\textbf {\bibinfo {volume} {46}},\ \bibinfo {pages} {313001} (\bibinfo
  {year} {2013})}\BibitemShut {NoStop}%
\bibitem [{\citenamefont {Schuwalow}\ \emph {et~al.}(2019)\citenamefont
  {Schuwalow}, \citenamefont {Schroeter}, \citenamefont {Gukelberger},
  \citenamefont {Thomas}, \citenamefont {Strocov}, \citenamefont {Gamble},
  \citenamefont {Chikina}, \citenamefont {Caputo}, \citenamefont {Krieger},
  \citenamefont {Gardner} \emph {et~al.}}]{schuwalow2019band}%
  \BibitemOpen
  \bibfield  {author} {\bibinfo {author} {\bibfnamefont {S.}~\bibnamefont
  {Schuwalow}}, \bibinfo {author} {\bibfnamefont {N.}~\bibnamefont
  {Schroeter}}, \bibinfo {author} {\bibfnamefont {J.}~\bibnamefont
  {Gukelberger}}, \bibinfo {author} {\bibfnamefont {C.}~\bibnamefont {Thomas}},
  \bibinfo {author} {\bibfnamefont {V.}~\bibnamefont {Strocov}}, \bibinfo
  {author} {\bibfnamefont {J.}~\bibnamefont {Gamble}}, \bibinfo {author}
  {\bibfnamefont {A.}~\bibnamefont {Chikina}}, \bibinfo {author} {\bibfnamefont
  {M.}~\bibnamefont {Caputo}}, \bibinfo {author} {\bibfnamefont
  {J.}~\bibnamefont {Krieger}}, \bibinfo {author} {\bibfnamefont {G.~C.}\
  \bibnamefont {Gardner}},  \emph {et~al.},\ }\href@noop {} {\bibfield
  {journal} {\bibinfo  {journal} {arXiv preprint arXiv:1910.02735}\ } (\bibinfo
  {year} {2019})}\BibitemShut {NoStop}%
\bibitem [{\citenamefont {van Weperen}\ \emph {et~al.}(2013)\citenamefont {van
  Weperen}, \citenamefont {Plissard}, \citenamefont {Bakkers}, \citenamefont
  {Frolov},\ and\ \citenamefont {Kouwenhoven}}]{van2013quantized}%
  \BibitemOpen
  \bibfield  {author} {\bibinfo {author} {\bibfnamefont {I.}~\bibnamefont {van
  Weperen}}, \bibinfo {author} {\bibfnamefont {S.~R.}\ \bibnamefont
  {Plissard}}, \bibinfo {author} {\bibfnamefont {E.~P.}\ \bibnamefont
  {Bakkers}}, \bibinfo {author} {\bibfnamefont {S.~M.}\ \bibnamefont {Frolov}},
  \ and\ \bibinfo {author} {\bibfnamefont {L.~P.}\ \bibnamefont
  {Kouwenhoven}},\ }\href@noop {} {\bibfield  {journal} {\bibinfo  {journal}
  {Nano letters}\ }\textbf {\bibinfo {volume} {13}},\ \bibinfo {pages} {387}
  (\bibinfo {year} {2013})}\BibitemShut {NoStop}%
\bibitem [{\citenamefont {Doh}\ \emph {et~al.}(2005)\citenamefont {Doh},
  \citenamefont {van Dam}, \citenamefont {Roest}, \citenamefont {Bakkers},
  \citenamefont {Kouwenhoven},\ and\ \citenamefont
  {De~Franceschi}}]{doh2005tunable}%
  \BibitemOpen
  \bibfield  {author} {\bibinfo {author} {\bibfnamefont {Y.-J.}\ \bibnamefont
  {Doh}}, \bibinfo {author} {\bibfnamefont {J.~A.}\ \bibnamefont {van Dam}},
  \bibinfo {author} {\bibfnamefont {A.~L.}\ \bibnamefont {Roest}}, \bibinfo
  {author} {\bibfnamefont {E.~P.}\ \bibnamefont {Bakkers}}, \bibinfo {author}
  {\bibfnamefont {L.~P.}\ \bibnamefont {Kouwenhoven}}, \ and\ \bibinfo {author}
  {\bibfnamefont {S.}~\bibnamefont {De~Franceschi}},\ }\href@noop {} {\bibfield
   {journal} {\bibinfo  {journal} {science}\ }\textbf {\bibinfo {volume}
  {309}},\ \bibinfo {pages} {272} (\bibinfo {year} {2005})}\BibitemShut
  {NoStop}%
\bibitem [{\citenamefont {Courtois}\ \emph {et~al.}(2008)\citenamefont
  {Courtois}, \citenamefont {Meschke}, \citenamefont {Peltonen},\ and\
  \citenamefont {Pekola}}]{courtois2008origin}%
  \BibitemOpen
  \bibfield  {author} {\bibinfo {author} {\bibfnamefont {H.}~\bibnamefont
  {Courtois}}, \bibinfo {author} {\bibfnamefont {M.}~\bibnamefont {Meschke}},
  \bibinfo {author} {\bibfnamefont {J.}~\bibnamefont {Peltonen}}, \ and\
  \bibinfo {author} {\bibfnamefont {J.~P.}\ \bibnamefont {Pekola}},\
  }\href@noop {} {\bibfield  {journal} {\bibinfo  {journal} {Physical Review
  Letters}\ }\textbf {\bibinfo {volume} {101}},\ \bibinfo {pages} {067002}
  (\bibinfo {year} {2008})}\BibitemShut {NoStop}%
\bibitem [{\citenamefont {Kulik}\ and\ \citenamefont
  {Omel'Yanchuk}(1977)}]{kulik1977properties}%
  \BibitemOpen
  \bibfield  {author} {\bibinfo {author} {\bibfnamefont {I.}~\bibnamefont
  {Kulik}}\ and\ \bibinfo {author} {\bibfnamefont {A.}~\bibnamefont
  {Omel'Yanchuk}},\ }\href@noop {} {\bibfield  {journal} {\bibinfo  {journal}
  {Sov. J. Low Temp. Phys.(Engl. Transl.);(United States)}\ }\textbf {\bibinfo
  {volume} {3}} (\bibinfo {year} {1977})}\BibitemShut {NoStop}%
\bibitem [{\citenamefont {Xiang}\ \emph {et~al.}(2006)\citenamefont {Xiang},
  \citenamefont {Vidan}, \citenamefont {Tinkham}, \citenamefont {Westervelt},\
  and\ \citenamefont {Lieber}}]{xiang2006ge}%
  \BibitemOpen
  \bibfield  {author} {\bibinfo {author} {\bibfnamefont {J.}~\bibnamefont
  {Xiang}}, \bibinfo {author} {\bibfnamefont {A.}~\bibnamefont {Vidan}},
  \bibinfo {author} {\bibfnamefont {M.}~\bibnamefont {Tinkham}}, \bibinfo
  {author} {\bibfnamefont {R.~M.}\ \bibnamefont {Westervelt}}, \ and\ \bibinfo
  {author} {\bibfnamefont {C.~M.}\ \bibnamefont {Lieber}},\ }\href@noop {}
  {\bibfield  {journal} {\bibinfo  {journal} {Nature Nanotechnology}\ }\textbf
  {\bibinfo {volume} {1}},\ \bibinfo {pages} {208} (\bibinfo {year}
  {2006})}\BibitemShut {NoStop}%
\bibitem [{\citenamefont {Nilsson}\ \emph {et~al.}(2012)\citenamefont
  {Nilsson}, \citenamefont {Samuelsson}, \citenamefont {Caroff},\ and\
  \citenamefont {Xu}}]{nilsson2012supercurrent}%
  \BibitemOpen
  \bibfield  {author} {\bibinfo {author} {\bibfnamefont {H.}~\bibnamefont
  {Nilsson}}, \bibinfo {author} {\bibfnamefont {P.}~\bibnamefont {Samuelsson}},
  \bibinfo {author} {\bibfnamefont {P.}~\bibnamefont {Caroff}}, \ and\ \bibinfo
  {author} {\bibfnamefont {H.}~\bibnamefont {Xu}},\ }\href@noop {} {\bibfield
  {journal} {\bibinfo  {journal} {Nano Letters}\ }\textbf {\bibinfo {volume}
  {12}},\ \bibinfo {pages} {228} (\bibinfo {year} {2012})}\BibitemShut
  {NoStop}%
\bibitem [{\citenamefont {Gharavi}\ \emph {et~al.}(2017)\citenamefont
  {Gharavi}, \citenamefont {Holloway}, \citenamefont {LaPierre},\ and\
  \citenamefont {Baugh}}]{gharavi2017nb}%
  \BibitemOpen
  \bibfield  {author} {\bibinfo {author} {\bibfnamefont {K.}~\bibnamefont
  {Gharavi}}, \bibinfo {author} {\bibfnamefont {G.~W.}\ \bibnamefont
  {Holloway}}, \bibinfo {author} {\bibfnamefont {R.~R.}\ \bibnamefont
  {LaPierre}}, \ and\ \bibinfo {author} {\bibfnamefont {J.}~\bibnamefont
  {Baugh}},\ }\href@noop {} {\bibfield  {journal} {\bibinfo  {journal}
  {Nanotechnology}\ }\textbf {\bibinfo {volume} {28}},\ \bibinfo {pages}
  {085202} (\bibinfo {year} {2017})}\BibitemShut {NoStop}%
\end{thebibliography}%

\end{document}